\documentclass[superscriptaddress,aps,prd,showpacs,preprint]{revtex4-1}
\usepackage{lingmacros}
\usepackage{tree-dvips}
\usepackage{graphicx}
\usepackage{subfigure}
\usepackage{amsmath}

\usepackage{latexsym}
\usepackage{color}
\usepackage{fullpage}
\usepackage{commath}
\usepackage{array}
\usepackage{dcolumn}
\usepackage[utf8]{inputenc}
\usepackage{graphicx, psfrag}
\usepackage{float}
\usepackage{tensor}
\usepackage{amsfonts}
\usepackage{hyperref}
\usepackage{pgfplots}
\usepackage{epstopdf}
\usepackage{booktabs}
\usepackage{comment}
\usepackage{mathtools}
\usepackage[dvipsnames]{xcolor}
\usepackage{mathrsfs}
\usepackage[normalem]{ulem}
\usepackage{cancel}

\newcommand{\VH}{V_{\mathbb{H}^{d-1}}}

\begin{document}

\title{Extended Thermodynamics and R\'enyi Entropy Beyond Fixed Central Charge}

\author{Chatchai Promsiri} 
\email{chatchaipromsiri@gmail.com} 
\affiliation{Quantum Computing and Information Research Centre (QX), Department of Physics, Faculty of Science, King Mongkut's University of Technology Thonburi,  Bangkok 10140, Thailand}

\author{Phuwadon Chunaksorn} 
\email{maxwelltle@gmail.com}
\affiliation{The Institute for Fundamental Study, Naresuan University, Phitsanulok, 65000, Thailand}

\author{Ratchaphat Nakarachinda}
\email{ratchaphat.n@rumail.ru.ac.th}
\affiliation{Quantum and Gravity Theory Research Group, Department of Physics, \\Faculty of Science, Ramkhamhaeng University, 282 Ramkhamhaeng Road,\\ Hua mak, Bang Kapi, Bangkok 10240, Thailand}

\author{Ekapong Hirunsirisawat}
\email{ekapong.hir@kmutt.ac.th}
\affiliation{Quantum Computing and Information Research Centre (QX), Department of Physics, Faculty of Science, King Mongkut's University of Technology Thonburi, Bangkok 10140, Thailand}

\begin{abstract}
An outstanding problem in the framework of conformal thermodynamics concerns the interpretation of variations in the central charge $C$. 
In this paper, we construct a novel central-charge R\'enyi entropy via the Casini-Huerta-Myers (CHM) map by considering thermal CFTs on a hyperbolic cylinder within a fixed charge, field theory volume and central charge potential  $(\tilde{Q},\mathcal{V},\mu_C)$ grand canonical ensemble.
We demonstrate that the resulting entropy satisfies all four fundamental R\'enyi entropy inequalities throughout the admissible range of $\mu_C$, establishing its consistency as a genuine R\'enyi measure.
Physically, this novel measure extends conventional R\'enyi entropy by capturing the degree of entanglement across a statistical ensemble of holographic CFTs with fluctuating degrees of freedom.
Furthermore, our conformal thermodynamic analysis of near-extremal configurations reveals that residual entropy arises from the central charge sector rather than thermal excitations. 
The mass gap that separates the extremal state and the first thermal excitation introduces a characteristic temperature scale $\tilde{T}_*$, which translates via the CHM map into a distinguished characteristic R\'enyi index $n_*$. Crucially, we propose that $n_*$ separates the theory space into two qualitatively distinct statistical regimes: a dominant-theory regime ($n > n_*$) governed by the most probable CFT realizations, and a multi-theory regime ($n < n_*$) where a broader spectrum of fluctuating theories and higher-energy modular excitations becomes increasingly relevant.

\end{abstract}

\maketitle

\section{Introduction} \label{sec: intro}
\label{sec: intro}

The AdS/CFT correspondence, realized within asymptotically anti-de Sitter (AdS) spacetimes, posits a duality between weakly coupled bulk gravity and strongly coupled boundary gauge theories~\cite{Maldacena:1997re,Witten:1998qj,Gubser:1998bc}.
In this way, the gravitational description has proven highly effective for modeling strongly correlated condensed matter systems ~\cite{Hartnoll:2009sz} and exploring quantum information-theoretic quantities such as holographic entanglement entropy (HEE)~\cite{Ryu:2006bv}.
A notable example is the Casini-Huerta-Myers (CHM) map~\cite{Casini:2011kv}, which relates the entanglement entropy (EE) across a spherical region of radius $R$ in Minkowski space to the thermal entropy of a CFT at temperature $T_0=1/2\pi R$ on the hyperbolic cylinder $\mathbb{R}\times\mathbb{H}^{d-1}$. 
Through the AdS/CFT correspondence, the R\'enyi entropy can be computed from the thermodynamics of hyperbolic AdS black holes~\cite{Hung:2011nu}.

Recently, the framework of \textit{conformal thermodynamics} extends thermodynamic phase space in the boundary theories by promoting the central charge $C$ to a thermodynamic variable, while its conjugate variable is identified as the central charge potential $\mu_C$ \cite{Visser:2021eqk,PhysRevLett.130.181401,Cong:2021jgb,Qu:2022nrt,Bai:2022vmx,Promsiri:2024hrl}. 
Within this framework, the Smarr relation of the bulk AdS black hole is mapped to the Euler relation of the dual large-$N$ gauge theories. 
In particular, Ref.~\cite{Cong:2021jgb} investigated thermal behaviors of charged CFT dual to spherical RN-AdS black holes in the fixed $(\tilde{Q},\mathcal{V},\mu_C)$ ensemble and interpreted variations of $C$ as moving among a family of holographic CFTs with different numbers of degrees of freedom.
Motivated by the central role of black hole (BH) thermodynamics in holography, this raises the question of whether we can also apply the new extended phase space approach to examine the consequence of varying central charge in other areas of holography?

In this paper, we address this question through the holographic R\'enyi entropy. 
The CHM map allows the R\'enyi entropy to be expressed in terms of the free energies of the dual CFT evaluated at temperatures $T_0$ and $T_0/n$, where $n$ denotes the R\'enyi index. 
This thermal description provides a natural setting in which a new pair of conjugate variables $\{C,\mu_C\}$ introduced in conformal thermodynamics offers an additional probe of the entanglement structure of holographic CFTs.

We begin in Sec.~\ref{sec:CFT thermo} with a brief review of conformal thermodynamics and its holographic interpretation.
Since the thermodynamics of CFTs on the hyperbolic cylinder play a central role in the holographic calculation of the R\'enyi entropy, in Sec.~\ref{sec:thermal CFT} we investigate the thermal behaviors of CFTs, which are dual to hyperbolic RN-AdS BHs within the conformal thermodynamic framework, considering both the fixed $(\tilde{\mu}_Q,\mathcal{V},C)$ and $(\tilde{Q},\mathcal{V},\mu_C)$ ensembles.

However, several conceptual issues concerning hyperbolic BHs remain open. 
For instance, the massless configuration exhibits a non-zero temperature and entropy due to the finite horizon radius.
By contrast, the extremal configuration with negative mass has a vanishing temperature but carries a residual entropy.
Understanding the origin and thermodynamic interpretation for both entropy remains an outstanding problem~\cite{Emparan:1999gf,Cai:2004pz}.
Crucially, this non-trivial entropy challenges the standard thermodynamic assumptions of extensivity and homogeneity. 
It signals a breakdown of the Euler relation in dual field theory unless additional thermodynamic variables, such as those appearing in an extended phase space, are properly taken into account.

Furthermore, the statistical description of black hole's thermodynamics may break down as one approaches the extremal limit due to uncontrollable fluctuations, as argued in early studies \cite{Preskill:1991tb}.
This breakdown is the existence of a \textit{mass gap}, which is defined as a finite energy separation between the extremal state and the lowest-lying excitations.
This gap introduces a characteristic temperature scale $T_*$.
Thus, the thermal behavior of low-temperature $T<T_*$ is entirely controlled by the properties of the extremal state, making the interpretation of the residual entropy particularly subtle.

Motivated by these considerations, in Sec.~\ref{sec:Residual Entropy}, we revisit the thermodynamics of the field theory, which is holographically dual to hyperbolic RN-AdS BH in the massless and extremal configurations, employing the conformal thermodynamics approach.
Rather than addressing the microscopic origin of the mass gap itself \cite{Emparan:1999gf,Iliesiu:2020qvm}, we examine how the extended thermodynamic variables modify the Euler relation and provide a consistent thermodynamic interpretation of these distinguished configurations.

Since the CHM map establishes a connection between the thermal CFTs on hyperbolic space and the entanglement structure in Minkowski space, it provides a framework for exploring the holographic interpretation of the characteristic temperature scale $T_*$ associated with the mass gap. 
Different limits of the R\'enyi index $n$ probe distinct features of the reduced density matrix $\rho_A$.
In the limit $n \to \infty$, the entropy is dominated by the largest eigenvalue $\lambda_{\max}$, corresponding via the CHM map to the low-temperature regime.
In contrast, the limit $n \to 0$ captures the logarithm of the rank of $\rho_A$, associated with the high-temperature regime.
These two limits are known as the Min-entropy and Hartley entropy, respectively.
This observation suggests the existence of a R\'enyi index that serves as a diagnostic tool for interpolating between the Min-entropy and Hartley entropy.
In Sec.~\ref{sec:index}, we demonstrate that the characteristic temperature scale $T_*$ gives rise to a distinguished R\'enyi index $n_*$ through the thermal description of holographic R\'enyi entropy.
In some sense, $n_*$ analogous to the role of the mass gap scale separating the extremal state from the lowest-lying exitations in the bulk black hole geometry. 
We then investigate the physical implications of this index through the charged R\'enyi entropy in the fixed $(\tilde{\mu}_Q,\mathcal{V},C)$ ensemble.

Our analysis in Sec.~\ref{sec:centralRenyi} is performed in the fixed $(\tilde{Q}, \mathcal{V}, \mu_C)$ ensemble. 
Within this framework, varying the R\'enyi index $n$ changes $T_0$ to $T_0/n$ of the thermal CFT through the CHM map, which induces nontrivial responses of the thermodynamic variables, including the central charge $C$. 
This allows us to examine how the entanglement structure evolves across a family of holographic CFTs with different numbers of degrees of freedom. In particular, we introduce a \textit{central-charge R\'enyi entropy} as a new information-theoretic probe that captures the response of the R\'enyi entropy under variations of $C$, providing a connection between the entanglement measure and the thermodynamic role of the central charge.
We conclude our study in Sec.~\ref{sec:concl} with a discussion and some comments on future work.

\section{Conformal Thermodynamic Approach}\label{sec:CFT thermo}

In this section, we provide a brief overview of the conformal thermodynamics approach and examine the thermodynamics of the conformal invariant system that is dual to a hyperbolic AdS-BH.

The spacetime metric for $d-$dimensional CFTs dual to the Einstein gravity in $(d+1)-$dimensions is
\begin{eqnarray}
    ds_\text{CFT}^2=\omega^2(-dt^2+L^2ds_{K,d-1}^2)=-\omega^2dt^2+(\omega L)^2ds_{K,d-1}^2, \label{cf metric}
\end{eqnarray}
where $ds_{K,d-1}^2$ denotes the line element for a unit $(d-1)-$dimensional hyperbolic, planar and sphere corresponding to the constant of curvature $K=-1,0$ and $1$, respectively.
An arbitrary dimensionless $\omega$ is the conformal factor that reflects the conformal symmetry of boundary theory.
The dual CFT has a central charge $C_d$ that relates to the gravitational parameters in the bulk as follows
\begin{eqnarray}
    C_d= N^{d/2}\sim \frac{L^{d-1}}{G_\text{N}}, \label{Central}
\end{eqnarray}
where $N$ denotes the rank of the gauge group \cite{Emparan:1999gf,Karch:2015rpa,Bai:2022vmx}.   
In $\text{AdS}_3/\text{CFT}_2$ correspondence, the Brown-Henneaux dictionary gives the central charge of the $\text{CFT}_2$ as $c = 3L/(2G_\text{N})$ \cite{Brown:1986nw}.
The central charges for dual CFTs in $d=3$ and $4$ are $\displaystyle C_3=N^{3/2}=3L^2/(2\sqrt{2}G_\text{N})$ and $C_4= N^2=\pi L^3/(2G_\text{N})$, respectively.

From the metric in Eq.~\eqref{cf metric}, the field theory volume is given by $\mathcal{V}\sim (\omega L)^{d-1}$.
For $\omega=1$, as in the conventional case, the AdS radius $L$ matches the boundary curvature radius $R$, which implies that a variation of $L$ in the bulk induces corresponding changes in both $\mathcal{V}$ and $C_d$ on the field theory side.
However, if one chooses the conformal factor $\omega=R/L$, the field theory volume and central charge are independent, namely $\mathcal{V}\sim R^{d-1}$ and $C_d\sim L^{d-1}/{G_\text{N}}$, respectively.
This allows us to study a holographic bulk-boundary first law by varying $\Lambda$ while keeping $G_\text{N}$ fixed.

Consider the $(d+1)-$dimensional Einstein-Maxwell theory with negative cosmological constant 
\begin{eqnarray}
S=\frac{1}{16\pi G_\text{N}}\int d^{d+1}x\sqrt{-g}\left( \mathcal{R}+\frac{d(d-1)}{L^2}-\frac{\ell_*^2}{4}F_{\mu \nu}F^{\mu \nu} \right), \label{EM action}
\end{eqnarray}
where $g$ and $\mathcal{R}$ are the determinant of the metric tensor and the Ricci scalar, respectively.
The constant $\ell_*$ describes the minimal coupling between gravity and the $U(1)$ gauge field $A_\mu$, where $F_{\mu\nu}=\partial_\mu A_\nu-\partial_\nu A_\mu$ is the Maxwell stress tensor.
Treating $\omega$ as thermodynamic parameter, the first law of BH thermodynamics in the bulk is extended as follows~\cite{PhysRevLett.130.181401}
\begin{eqnarray}
d\left(\frac{M}{\omega}\right)&=&\frac{T}{\omega}d\left(\frac{A_H}{4G_\text{N}}\right)+\frac{\Phi}{\omega L} d(QL)-\frac{M}{(d-1)\omega}\frac{d(\omega L)^{d-1}}{(\omega L)^{d-1}}\nonumber \\
&& \hspace{2.3cm}+\left( \frac{M}{\omega}-\frac{TS_\text{th}}{\omega}-\frac{\Phi Q}{ \omega}\right)\frac{d(L^{d-1}/G_\text{N})}{L^{d-1}/G_\text{N}}, \label{extended 1st law}
\end{eqnarray}
where $M, A_H, S_\text{th}$, and $T$ stand for the mass of BH, the area at the event horizon, the thermal entropy, and the Hawking temperature, respectively.
The holographic map between the bulk quantities (without tildes) and the dual CFT ones (with tildes) is given by
\begin{align}
    \tilde{S}_\text{th}=S_\text{th}=\frac{A_H}{4G_\text{N}}, \qquad
    \tilde{E}=\frac{M}{\omega}, \qquad
    \tilde{T}=\frac{T}{\omega}, \qquad
    \tilde{\Phi}=\frac{\Phi}{\omega L}, \qquad
    \tilde{Q}=QL.\label{map}
\end{align}
Defining the central charge
\begin{align}
    C=\frac{V_{K,d-1}}{16\pi}C_d
    =\frac{V_{K,d-1}L^{d-1}}{16\pi G_N},
\end{align}
where $V_{K,d-1}$ represents the volume of the unit $(d-1)-$dimensional space with constant curvature $K$.
The extended first law in the bulk in Eq.~\eqref{extended 1st law} provides the first law of the dual CFT as
\begin{eqnarray}
d\tilde{E}=\tilde{T}d\tilde{S}_\text{th}+\tilde{\Phi}d\tilde{Q}-pd\mathcal{V}+\mu_CdC,
\end{eqnarray}
where the conjugate potential of the central charge and the pressure for the CFT system are defined by
\begin{eqnarray}
    \mu_C=\frac{1}{C}\left(\tilde{E}-\tilde{T}\tilde{S}_\text{th}-\tilde{\Phi}\tilde{Q}\right),\qquad 
    p=\frac{\tilde{E}}{(d-1)\mathcal{V}}, \label{p1}
\end{eqnarray}
respectively. 
In this approach, the CFT volume can be written in terms of $\omega L$ as
\begin{align}
    \mathcal{V}=V_{K,d-1}(\omega L)^{d-1}.
\end{align}

\section{Thermodynamics of Dual CFT on Hyperbolic Space} \label{sec:thermal CFT}
To study a charged CFT on the boundary via holography, we introduce a $U(1)$ gauge field $A_{\mu}$ into the bulk action, which acts as a source for the conserved current operator $J^\mu$ in the boundary CFT. 
Specifically, the boundary value of the temporal component, $A_t \big|_{\rho \to \infty} = \mu_Q$, is identified with the chemical potential of the dual theory, while the subleading term in the asymptotic expansion of $A_t$ corresponds to the charge density $J^t$.
The corresponding field equations admit the RNAdS-BH solution with hyperbolic slicing as follows
\begin{eqnarray}
    ds^2 &=& -f(\rho)dt^2+\frac{d\rho^2}{f(\rho)}+\rho^2ds_{\mathbb{H}^{d-1}}^2,
\end{eqnarray}
with the horizon function
\begin{eqnarray}
    f(\rho)&=&\frac{\rho^2}{L^2}-1-\frac{m}{\rho^{d-2}}+\frac{q^2}{\rho^{2d-4}}. \label{f fn}
\end{eqnarray}
Here, $\rho$ is the radial coordinate and $m$ is the parameter related to the mass of BH.
The metric of the unit hyperbolic space $\mathbb{H}^{d-1}$ is given by
\begin{eqnarray}
    ds_{\mathbb{H}^{d-1}}^2=du^2+\sinh^2 u~ ds^2_{\mathbb{S}^{d-2}}. \label{unit hyp metric1}
\end{eqnarray}
The mass of black hole $M$ relates to the parameter $m$ via
\begin{eqnarray}
   M=\frac{(d-1)V_{\mathbb{H}^{d-1}}}{16\pi G_\text{N}}m=\frac{(d-1)V_{\mathbb{H}^{d-1}}}{16\pi G_\text{N}}\rho_+^{d-2}\left[\left( \frac{\rho_+}{L}\right)^2-1+\frac{q^2}{\rho_+^{2d-4}}\right], \label{charged BH mass} 
\end{eqnarray}
where $\VH$ represents an infinite volume of hyperbolic space.
By using $\displaystyle Q=\frac{\ell_*^2}{16\pi G_\text{N}}\oint*F$, where $*$ represents the Hodge dual, the electric charge $Q$ of BH is related to the parameter $q$ through
\begin{eqnarray}
    Q=\frac{(d-1)V_{\mathbb{H}^{d-1}}}{16\pi G_\text{N}}\eta \ell_*q, \ \ \ \text{with}\ \ \ \eta =\sqrt{\frac{2(d-2)}{d-1}}.
\end{eqnarray}
The bulk gauge field is given by
\begin{eqnarray}
    A=\left( \frac{2}{\eta}\frac{q}{\ell_*\rho^{d-2}}-\frac{\mu_Q}{2\pi }\right)dt, \label{gauge A}
\end{eqnarray}
where $\mu_Q$ is the electric potential.
By fixing the gauge field vanishes at the BH's horizon, the electric potential will be written as
\begin{eqnarray}
    \mu_Q =\frac{4\pi}{\eta}\frac{q}{\ell_*\rho_+^{d-2}}. \label{elec pro} 
\end{eqnarray}
The thermal entropy follows the area law, it reads
\begin{eqnarray}
    S_\text{th}=\frac{V_{\mathbb{H}^{d-1}}}{4G_\text{N}}\rho_+^{d-1}=\frac{V_{\mathbb{H}^{d-1}} L^{d-1}}{4G_\text{N}}x^{d-1}, \label{thermal entropy}
\end{eqnarray}
and Hawking temperature in terms of $q$ and $\mu_Q$ are 
\begin{eqnarray}
    T&=&\frac{f'(\rho_+)}{4\pi}=\frac{d-2}{4\pi Lx}\left( \frac{d}{d-2}x^2-1-\frac{q^2}{L^{2d-4}x^{2d-4}}\right), \label{temperature q}\\
    &=&\frac{d-2}{4\pi Lx}\left[ \frac{d}{d-2}x^2-1-\frac{d-2}{2(d-1)}\left( \frac{\mu_Q \ell_*}{2\pi}\right)^2 \right],
\end{eqnarray}
where $x=\rho_+/L$.
The last equality is obtained using Eq.~\eqref{elec pro}.
In contrast to the $k=0,+1$ classes of AdS-BH, the $k=-1$ class possesses a horizon at $\rho_+=L$ even in the massless limit $M=0$ and $Q=0$. 
The Hawking temperature associated with this horizon is given by
\begin{align}
    T_0=\frac{1}{2\pi L}.
\end{align}
Moreover, there exists a range of negative mass values for which the BH still possesses a regular event horizon.
The smallest possible mass corresponds to the extremal configuration, where the temperature vanishes and the horizon radius reaches its minimum value.
These extremal quantities are given by
\begin{align}
    m_\text{ext}=-\frac{2}{d-2}\left(\frac{d-2}{d}\right)^{d/2}L^{d-2}, \qquad \rho_\text{ext}=\sqrt{\frac{d-2}{d}}L.
\end{align}

Given the thermodynamic quantities defined above, the extended thermodynamic relations such as the Smarr formula and the first law within black hole chemistry can be constructed as \cite{Kubiznak:2016qmn}
\begin{eqnarray}
    M&=&\left(\frac{d-1}{d-2}\right)TS_\text{th}+\frac{\mu_Q}{2\pi}Q-\left(\frac{2}{d-2}\right)PV, \label{Smarr eps} \\
    dM&=&TdS_\text{th}+\frac{\mu_Q}{2\pi}dQ+VdP. \label{1st law eps}
\end{eqnarray}
(see App.~\ref{App A} for a detailed derivation).
Next, we will discuss how the CFT thermodynamics can be obtained using the map~\eqref{map} from the BH ones.

The resulting holographic first law for CFT which holographically dual to hyperbolic RNAdS-BH reads
\begin{eqnarray}
d\tilde{E}=\tilde{T}d\tilde{S}_\text{th}+\frac{\tilde{\mu}_Q}{2\pi}d\tilde{Q}-pd\mathcal{V}+\mu_CdC,
\end{eqnarray}
with the Euler relation:
\begin{eqnarray}
\tilde{E}=\tilde{T}\tilde{S}_\text{th}+\frac{\tilde{\mu}_Q}{2\pi}\tilde{Q}+\mu_CC
=(d-1)p\mathcal{V}.\label{Euler cft}
\end{eqnarray}
In App.~\ref{App A}, we elaborate on the detailed calculations.
Here we consider the thermodynamics of dual CFT living in the $\mathbb{R}\times \mathbb{H}^{d-1}$ spacetime.
Using the dictionary in Eq.~\eqref{dictionary} and defining a dimensionless parameter $y=q/L^{d-2}$, we obtain the internal energy $\tilde{E}$, thermal entropy $\tilde{S}_\text{th}$ and temperature $\tilde{T}$ of dual CFT from Eqs.~\eqref{charged BH mass}, \eqref{thermal entropy} and \eqref{temperature q}, respectively, are found to be:
\begin{eqnarray}
    \tilde{E}&=&\frac{(d-1)C}{R}x^{d-2}\left( x^2-1+\frac{y^2}{x^{2d-4}}\right), \label{E CFT} \\
    \tilde{S}_\text{th}&=&4\pi Cx^{d-1}, \label{Sth} \\
    \tilde{T}&=&\frac{d-2}{4\pi R}\frac{1}{x}\left( \frac{d}{d-2}x^2-1-\frac{y^2}{x^{2d-4}}\right). \label{CFT TH}
\end{eqnarray}
Furthermore, the central charge potential $\mu_C$ and the field theory pressure $p$ are given by the first and second relations of Eq.~\eqref{p1}, yielding
\begin{align}
    \mu_C&=-\frac{1}{R}x^{d-2}\left(1+x^2+\frac{y^2}{x^{2d-4}}\right), \label{mu C} \\
    p&=\frac{C}{\VH R^d}x^{d-2}\left(x^2-1+\frac{y^2}{x^{2d-4}}\right).\label{p charged}
\end{align}
Here, we choose the conformal factor $\omega=R/L$. 
Consequently, the spatial volume of dual CFT is given by 
\begin{eqnarray}
    \mathcal{V}=\VH R^{d-1}.\label{V CFT}
\end{eqnarray}
We can map dimensionless bulk parameter $y$ to charge $\tilde{Q}$ and chemical potential $\tilde{\mu}_Q$ for dual CFT via the relations
\begin{eqnarray}
    \tilde{Q}=QL=(d-1)C\eta\ell_*y, 
    \qquad
    \tilde{\mu}_Q=\frac{\mu_Q}{R}=\frac{1}{R}\frac{4\pi }{\eta \ell_*}\frac{y}{x^{d-2}}.  \label{charge and potential}
\end{eqnarray}
It is noteworthy that eight distinct thermodynamic ensembles arise from extending the standard thermodynamics of RNAdS-BH through a conformal thermodynamic approach. 
In this work, we focus on the thermodynamics of the dual CFT within two specific grand canonical ensembles: the fixed $(\tilde{\mu}_Q, \mathcal{V}, C)$ ensemble and the fixed $(\tilde{Q}, \mathcal{V}, \mu_C)$ ensemble.

\subsection{Fixed $(\tilde{\mu}_Q,\mathcal{V},C)$ Ensemble}
Starting from the Helmholtz free energy $\tilde{F}_{\tilde{Q},\mathcal{V},C} = \tilde{E} - \tilde{T}\tilde{S}_{\rm th}$ in the canonical ensemble or fixed $(\tilde{Q},\mathcal{V},C)$ ensemble, the transition to the grand canonical ensemble is performed via a Legendre transformation. 
For our gauge field definition in Eq.\eqref{gauge A}, the Gibbs free energy $\tilde{\Phi}_{\tilde{\mu}_Q,\mathcal{V},C}$ associated with fixed $(\tilde{\mu}_Q,\mathcal{V},C)$ is obtained by
\begin{align}
\tilde{\Phi}_{\tilde{\mu}_Q,\mathcal{V},C}&= \tilde{F}_{\tilde{Q},\mathcal{V},C} - \frac{\tilde{\mu}_Q}{2\pi} \tilde{Q}, \nonumber \\
&=\tilde{E} - \tilde{T}\tilde{S}_\text{th}-\frac{\tilde{\mu}_Q}{2\pi} \tilde{Q}, \nonumber \\
&=\mu_CC. \label{Gibbs 1st}
\end{align}
Here, we used the Euler relation in the final step.
In the following, for simplicity, we denote this free energy as $\tilde{\Phi}_1 \equiv \tilde{\Phi}_{\tilde{\mu}_Q, \mathcal{V}, C}$.
The differential $d\tilde{\Phi}_1$ reads
\begin{align}
d\tilde{\Phi}_1 = -\tilde{S}d\tilde{T} - \frac{\tilde{Q}}{2\pi} d\tilde{\mu}_Q - p d\mathcal{V} + \mu_{C} dC.
\end{align}
Therefore, we express all thermodynamic quantities as functions of $\tilde{\mu}_Q, R,C$ and $x$ to study the thermal behavior of the dual CFT in this ensemble.
By eliminating $y$ from Eq.\eqref{CFT TH} using Eq.\eqref{charge and potential}, the Hawking temperature becomes
\begin{align}
    \tilde{T}&=\frac{d-2}{4\pi Rx}\left( \frac{d}{d-2}x^2-\Delta \right), \label{T Myers}
\end{align}
where $\Delta$ is defined as
\begin{align}
    \Delta=1+\frac{d-2}{2(d-1)}\left( \frac{\tilde{\mu}_Q R\ell_*}{2\pi}\right)^2. \label{def delta}
\end{align}
One can solve parameter $x$ in terms of $\tilde{T}$ and $\tilde{\mu}_Q$ as follows
\begin{align}
    x&=\frac{2\pi R\tilde{T}}{d}+\sqrt{\frac{4\pi^2R^2\tilde{T}^2}{d^2}+\frac{d-2}{d}\Delta}. \label{x for 1st en}
\end{align}
The internal energy $\tilde{E}_1$ associated with this ensemble is
\begin{align}
    \tilde{E}_1=\tilde{E}-\frac{\tilde{\mu}_Q}{2\pi}\tilde{Q}=\frac{(d-1)C}{R} x^{d-2} \left( x^2 - \Delta \right),
\end{align}
where the mass of BH in the bulk $\tilde{E}$ reads
\begin{align}
    \tilde{E}&=\frac{(d-1)C}{R}x^{d-2}\left( x^2-2+\Delta\right).
\end{align}
The heat capacity can be obtain as follows
\begin{align}
    C_1=\left( \frac{\partial\tilde{E}_1}{\partial \tilde{T}}\right)_{\tilde{\mu}_Q,\mathcal{V},C}=4(d-1)\pi Cx^{d-1}\frac{\left[dx^2-(d-2)\Delta\right]}{\left[dx^2+(d-2)\Delta\right]}. \label{heat capa 1st}
\end{align}
From the central charge potential
\begin{align}
    \mu_C&=-\frac{1}{R}x^{d-2}\left(x^2+\Delta\right),
\end{align}
the Gibbs free energy as described in Eq.\eqref{Gibbs 1st} can be written as follows
\begin{align}
    \tilde{\Phi}_1=-\frac{C}{R}x^{d-2}\left(x^2+\Delta\right).
\end{align}
\begin{figure}[h]
    \centering
    \includegraphics[width = 5.1 cm]{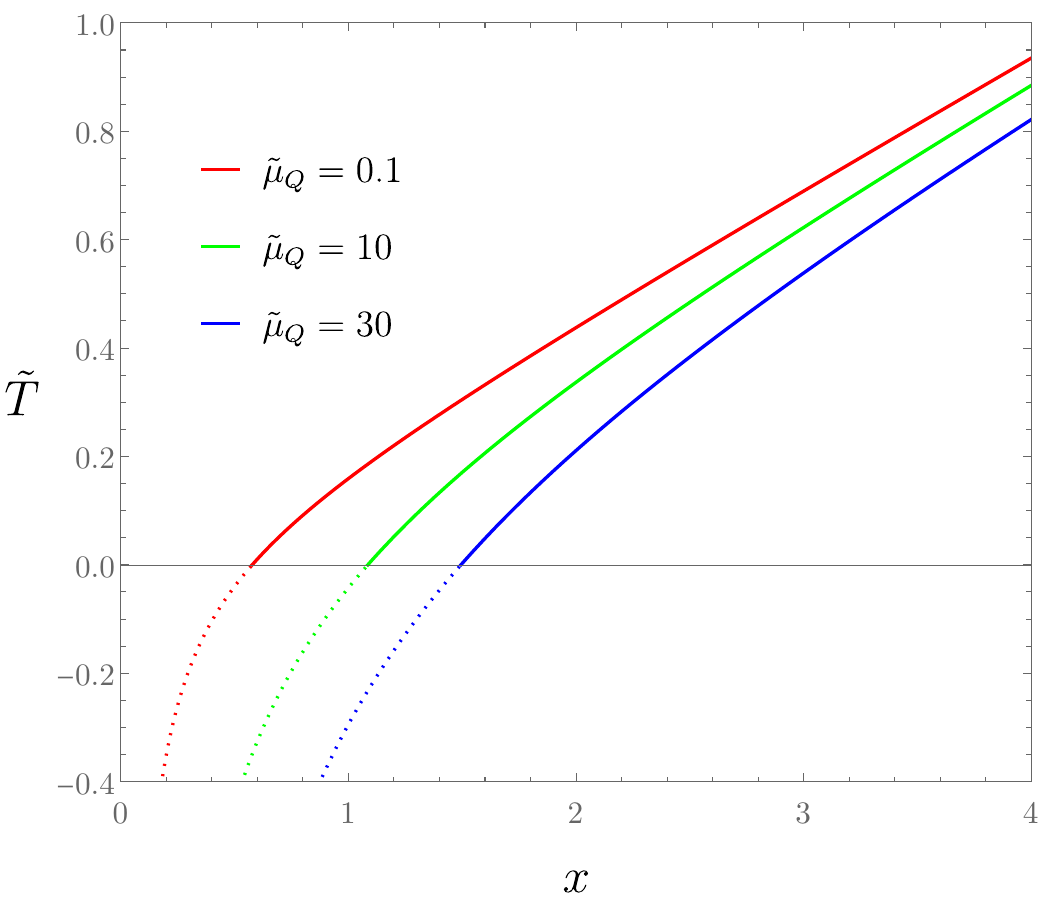} \hspace{0.3cm}
    \includegraphics[width = 5.2 cm]{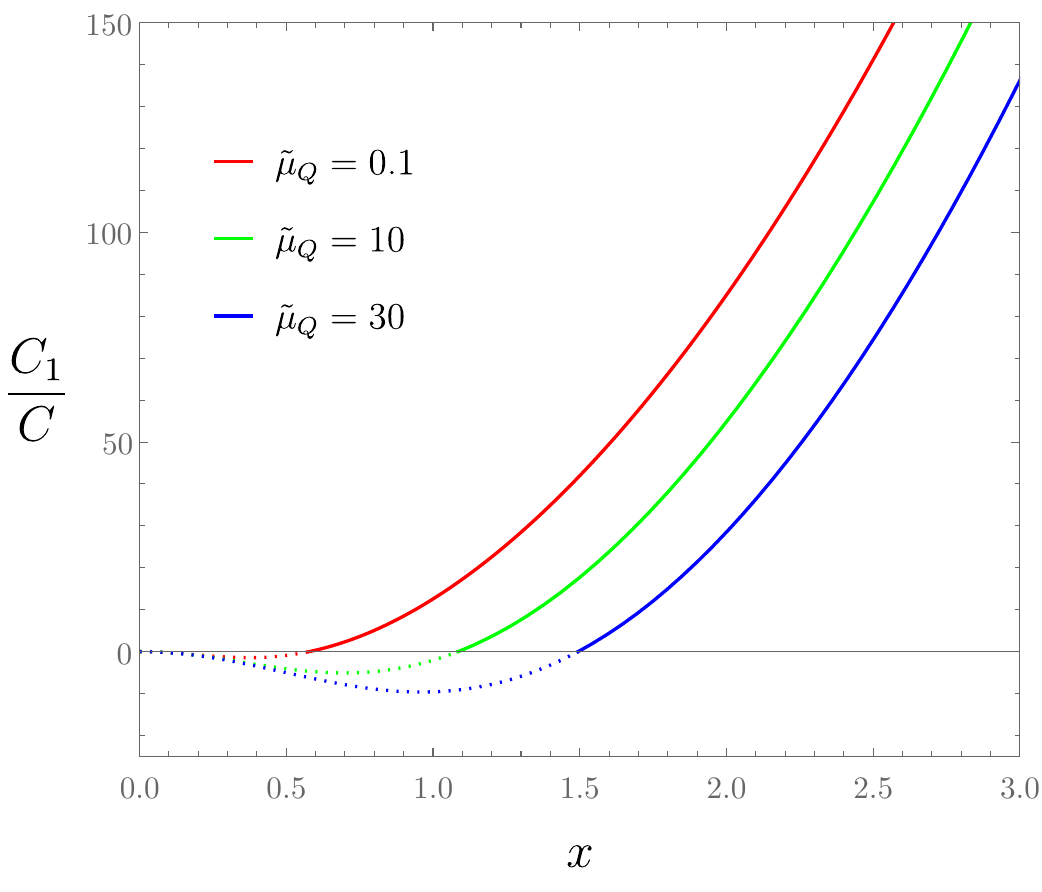}\hspace{0.3 cm}
    \includegraphics[width = 5.2 cm]{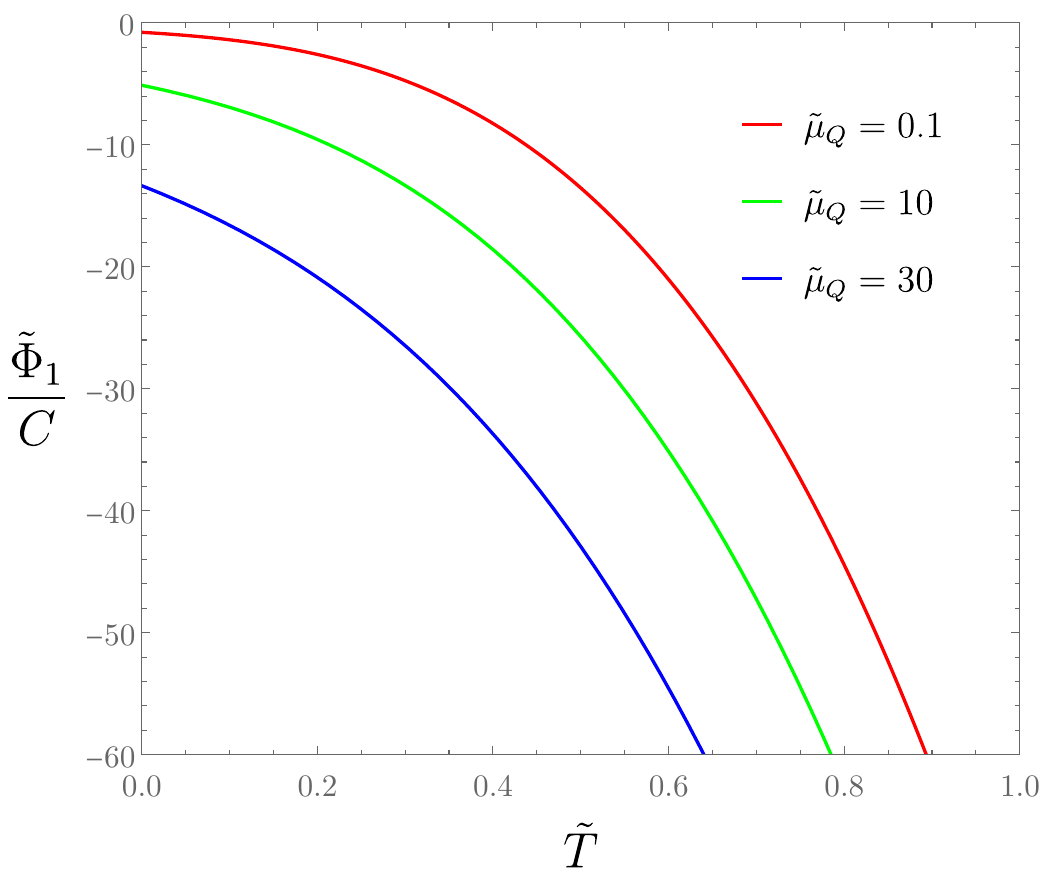}
\caption{The behaviors of temperature $\tilde{T}$ versus $x$ (left panel), the heat capacity $C_1/C$ versus $x$ (middle panel) and Gibbs free energy $\tilde{\Phi}_1/C$ versus temperature $\tilde{T}$ (right panel) of $3$-dimensional dual CFT in the $(\tilde{\mu}_Q,\mathcal{V},C)$ ensemble for various values of $\tilde{\mu}_Q$ with $R\,=\,\ell_*=1$. 
Note the dotted line correspond to the regime of $x<x_\text{min}$.} 
    \label{fig: 1st ensemble thermo}
\end{figure}

The extremal state for this ensemble is dependent on $\tilde{\mu}_Q$ and occurs at $x = x_{\text{min}}$, which is obtained by setting $\tilde{T}=0$ in Eq.\eqref{x for 1st en} with $\tilde{T}=0$:
\begin{align}
    x_\text{min}=\sqrt{\frac{d-2}{d}\Delta}\,.
\end{align}
The regime where $x < x_{\text{min}}$ indicates that black hole solutions do not exist in the bulk; this unphysical region is represented by the dotted lines in Fig.\ref{fig: 1st ensemble thermo}. In the physical domain where $x \geq x_{\text{min}}$, the temperature $\tilde{T}$ increases monotonically with $x$ (left panel).

The heat capacity $C_1/C$ (middle panel) remains positive throughout this range, indicate that these dual CFT states are thermodynamically stable. 
Furthermore, the Gibbs free energy $\tilde{\Phi}_1C/R$ (right panel) exhibits a smooth, single-branched behavior without a cusp, indicating that the dual CFT resides in a single stable phase for all temperatures $\tilde{T} > 0$.
Notably, while these numerical results are illustrated for $d=3$, the qualitative thermodynamic behavior remains identical for all dimensions $d \geq 3$.

\begin{figure}[h]
    \centering
     \includegraphics[width = 6.5 cm]{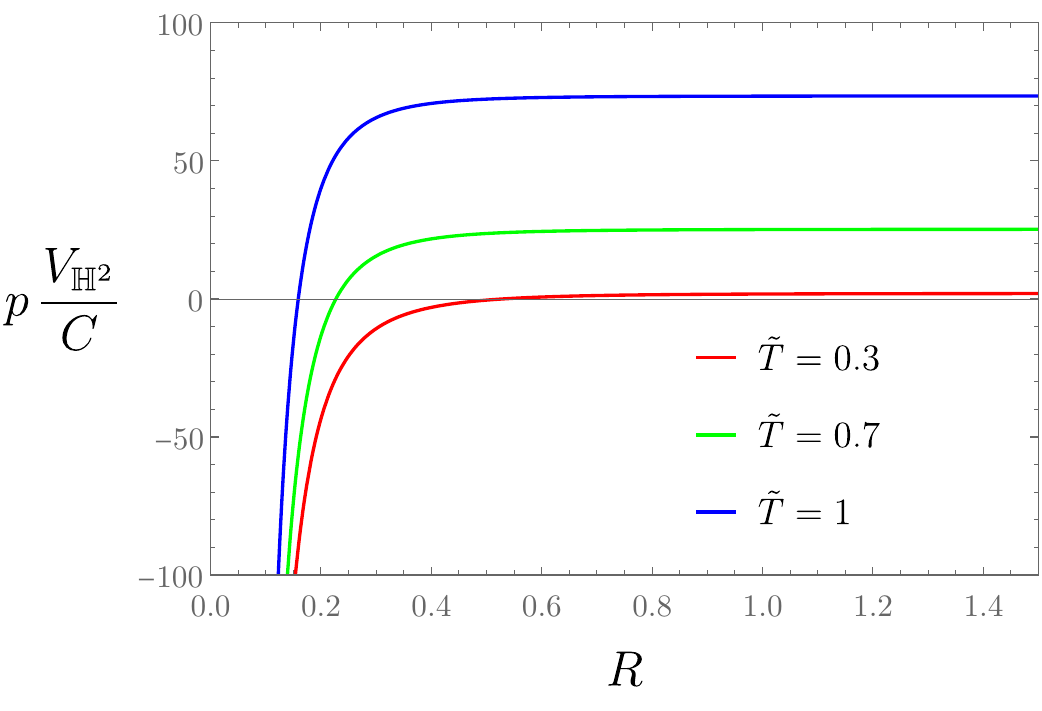}\qquad
     \includegraphics[width = 6.5 cm]{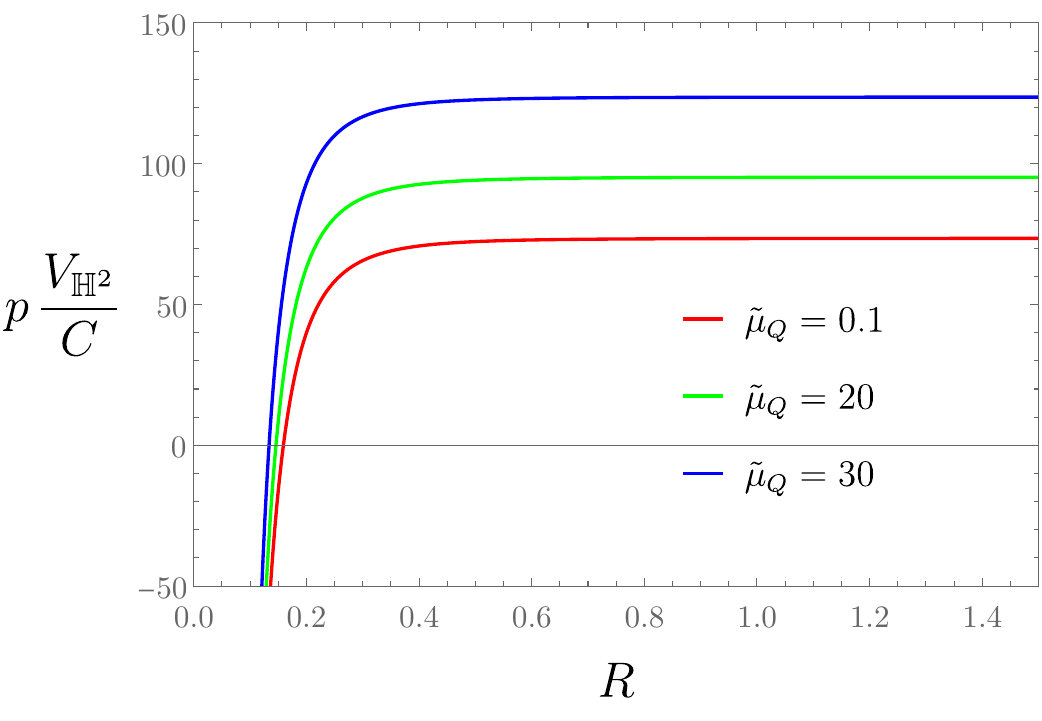}
    \caption{
    The pressure of the 3-dimensional dual CFT versus $R$ for various values of $\tilde{T}$ with $\tilde{\mu}_Q=1$ (left panel) and various values of $\tilde{\mu}_Q$ with $\tilde{T}=1$ (right panel) and $\ell_*$ is fixed as unity.} 
    \label{fig: P-R en1}
\end{figure}
\begin{figure}[h]
    \centering
     \includegraphics[width = 6.5 cm]{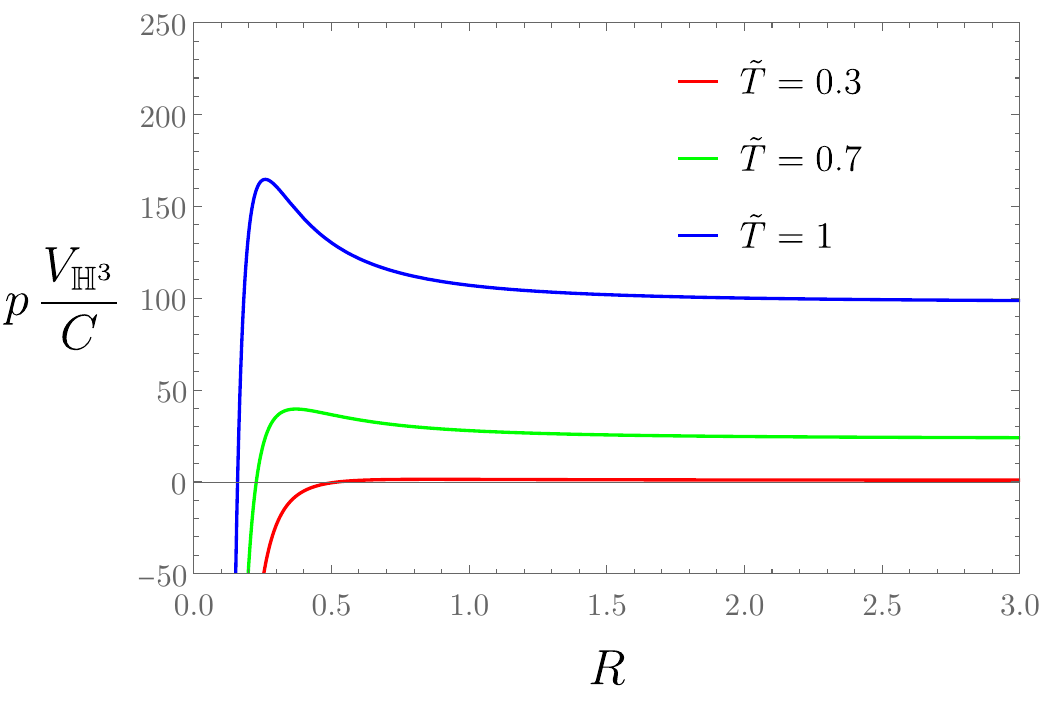}\qquad
     \includegraphics[width = 6.5 cm]{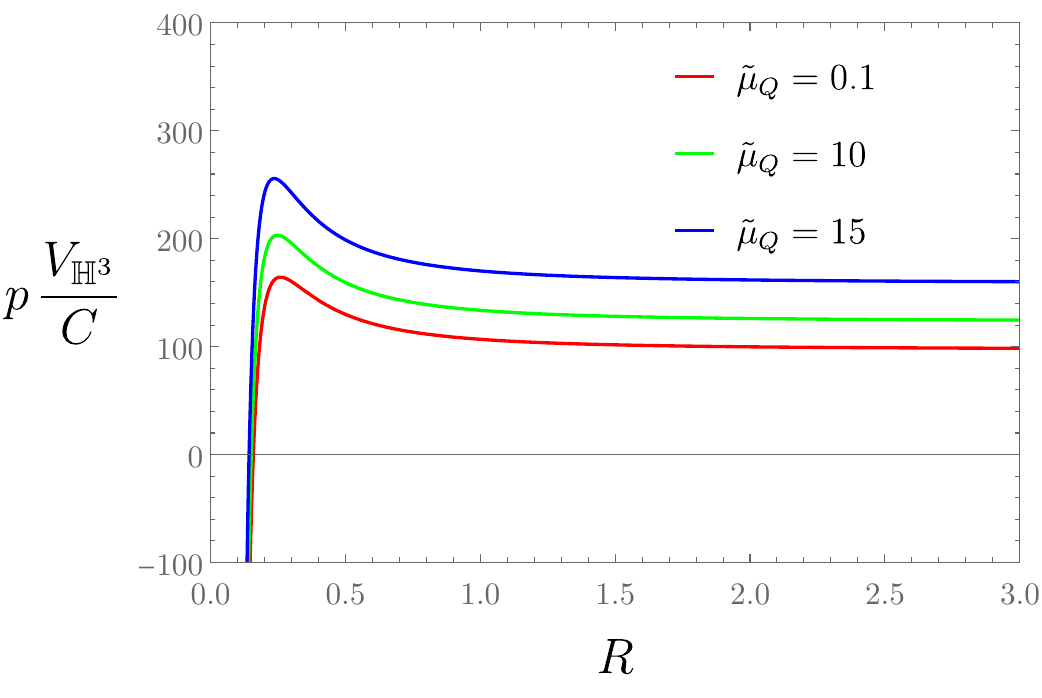}
    \caption{
    The pressure of the 4-dimensional dual CFT versus $R$ for various values of $\tilde{T}$ with $\tilde{\mu}_Q=1$ (left panel) and various values of $\tilde{\mu}_Q$ with $\tilde{T}=1$ (right panel) and $\ell_*$ is fixed as unity.} 
    \label{fig: P-R en1 d=4}
\end{figure}

The CFT pressure $p$ in this ensemble can be obtain from the equation of state in Eq.\eqref{p charged} as follows
\begin{align}
    p&=\frac{C}{\VH R^d}x^{d-2}\left( x^2-2+\Delta\right)
\end{align}
The behavior of CFT pressure $p$ as a function of finite size $R$ of the boundary with fixed temperature $\tilde{T}$ for $d=3$ is shown in Fig.~\ref{fig: P-R en1}. 
It is seen that the pressure increases monotonically in $R$ and saturates at some finite values at $R\to\infty$.
For sufficiently small $R$, the pressure can be negative.
By increasing both $\tilde{T}$ (with fixing $\tilde{\mu}_Q$) and $\tilde{\mu}_Q$ (with fixing $\tilde{T}$), the pressure will cross its zero value on smaller $R$.
For the case $d=4$, the behavior of the pressure shown in Fig.~\ref{fig: P-R en1 d=4} is significantly different from $d=3$.
We observe that a phase with negative slope emerges.
Equivalently, there exists the local maximum point in $p$,  indicating that $d=4$ serves as a critical dimension for this dual CFT, as integer dimensions $d \ge 4$ are required to support a mechanically stable phase  at large $R$.

\subsection{Fixed $(\tilde{Q}, \mathcal{V}, \mu_C)$ Ensemble}\label{sub 2nd ensemble}

Alternatively, by performing a Legendre transformation on $\tilde{F}_{\tilde{Q},\mathcal{V},C}$ with respect to the central charge $C$ instead of the charge $\tilde{Q}$, we obtain another grand canonical ensemble with fixed $(\tilde{Q}, \mathcal{V}, \mu_C)$. 
While the physical interpretation of this ensemble remains debated, varying the central charge should be viewed as moving across different holographic CFTs rather than changing the state within a fixed theory.
The corresponding thermodynamic potential, which we denote as the Gibbs free energy $\tilde{\Phi}_2$, is defined as
\begin{eqnarray}
\tilde{\Phi}_2=\tilde{E}-\tilde{T}\tilde{S}_\text{th}-\mu_CC=\frac{\tilde{\mu}_Q}{2\pi}\tilde{Q}. \label{F3}
\end{eqnarray}
The differential of $\tilde{\Phi}_2$ reads
\begin{eqnarray}
    d\tilde{\Phi}_2=-\tilde{S}_\text{th}d\tilde{T}+\frac{\tilde{\mu}_Q}{2\pi} d\tilde{Q}-pd\mathcal{V}-Cd\mu_C.
\end{eqnarray}
It is worthy to note that $\mu_C$ is always negative as shown in Eq.~\eqref{mu C}, in contrast to the spherical BH case where $\mu_C$ can take both positive and negative values \cite{Cong:2021jgb,Ladghami:2024wkv}.
To study the thermal behavior of the dual CFT in this ensemble, we express all thermodynamic quantities as functions of $\tilde{Q}, R,\mu_C$ and $x$.
For the temperature $\tilde{T}$ as described in Eq.~\eqref{CFT TH}, we eliminate term $y^2/x^{2d-4}$ by using Eq.~\eqref{mu C}, yielding
\begin{eqnarray}
    \tilde{T}&=&\frac{d-2}{4\pi R}\left[\frac{2(d-1)}{d-2}x+\frac{\mu_C R}{x^{d-1}} \right].\label{T3}  
\end{eqnarray}
In the limit of the $\tilde{E}=0$, where $\mu_C = -2/R$, the temperature in Eq.~\eqref{T3} reduces to $\tilde{T} = 1/(2\pi R)$.
The temperature is an increasing function of $x$.
However, the result is not physical for the whole range of $x$.
Our discussion of the validity range will be continued later.

It is important to note that the value of central charge is not fixed within this ensemble.
To express $C$ as a function of $\tilde{Q},\,R, \, \mu_C $ and $x$, we substitute $y= \tilde{Q}/\big[\ell_* C\sqrt{2(d-1)(d-2)}\,\big]$ from Eq.~\eqref{charge and potential} into Eq.~\eqref{mu C}, which gives
\begin{eqnarray}
    C=\frac{\tilde{Q}x^{2-\frac{d}{2}}}{\ell_*\sqrt{2(d-1)(d-2)}\sqrt{-x^d-x^{d+2}-\mu_C Rx^2}}.  \label{C ensemble}
\end{eqnarray}
Substituting $C$ in the above equation into Eq.~\eqref{Sth}, the thermal entropy is given by
\begin{eqnarray}
    \tilde{S}_\text{th}=\frac{2\sqrt{2}\pi \tilde{Q}x^{\frac{d}{2}+1}}{\ell_*\sqrt{(d-1)(d-2)}\sqrt{-x^d-x^{d+2}-\mu_C Rx^2}}. \label{S3}
\end{eqnarray}
The corresponding internal energy $\tilde{E}_2$ defined as $\tilde{E}_2=\tilde{E}-\mu_C C$, which can be expressed in terms of parameter $x$ by
\begin{align}
    \tilde{E}_2=-\frac{\tilde{Q}x^{-d/2}\left[2(d-1)x^d+d\,\mu_CRx^2\right]}{R\ell_*\sqrt{2(d-1)(d-2)}\sqrt{-x^d-x^{d+2}-\mu_CRx^2}}. \label{E2}
\end{align}
Consequently, the heat capacity can be computed via 
\begin{eqnarray}            
C_2=\left( \frac{\partial(\tilde{E}-\mu_CC)}{\partial \tilde{T}}\right)_{\tilde{Q},\mathcal{V},\mu_C}=\tilde{T}\left(\frac{\partial \tilde{S}_\text{th}}{\partial \tilde{T}} \right)_{\tilde{Q},\mathcal{V},\mu_C},
\end{eqnarray}
which can be explicitly expressed as follows
\begin{eqnarray}
C_2=\frac{\sqrt{2} \pi  \tilde{Q} x^{\frac{d}{2}+1} \left(-2 x^d-d R x^2 \mu _C\right) \big[-2 (d-1) x^d-(d-2) R \mu _C\big]}{(d-1)^{3/2} \sqrt{d-2} \ell_* \big[2 x^d-(d-2) R \mu _C\big] \left(-x^{d+2}-x^d-R x^2 \mu_C\right)^{3/2}}. \label{heat capa}
\end{eqnarray}
From the expression for the Gibbs free energy in Eq.~\eqref{F3}, i.e.,  $\tilde{\Phi}_2=\tilde{\mu}_Q \tilde{Q}/(2\pi)$, we write $\tilde{\mu}_Q$ in terms of $\mu_C$ using Eqs.~\eqref{mu C} and \eqref{charge and potential}, leading to
\begin{eqnarray}
    \tilde{\Phi}_2&=&\frac{2\tilde{Q}}{\eta \ell_*R}\frac{\sqrt{-x^d-x^{d+2}-\mu_CRx^2}}{x^{d/2}}. \label{F2 charge}
\end{eqnarray}

\begin{figure}[h]
    \centering
     \includegraphics[width = 6.5 cm]{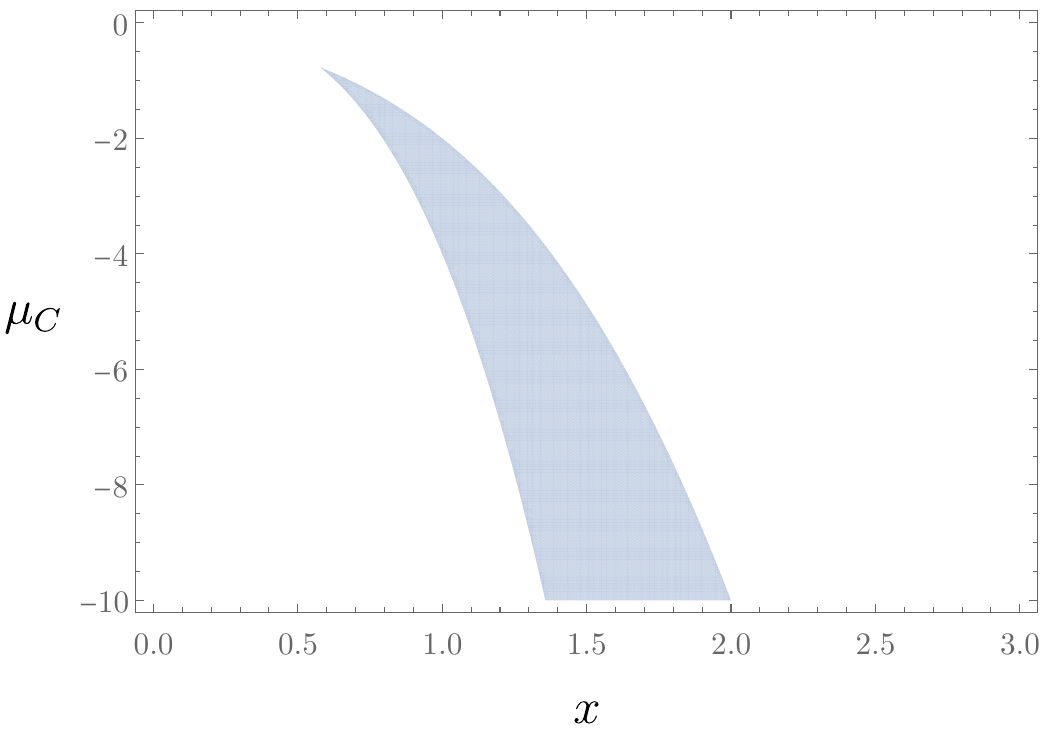}
    \caption{
    The region of $\mu_C$ and $x$ for the existence of thermal states in the dual CFT for $d=3$ and $R$ is fixed as unity.} 
    \label{fig: PS}
\end{figure}

Let us now examine the physical parameter space of this ensemble.
Given that $\tilde{T}\geq 0$ and $\tilde{\Phi}_2$ is a real number, the parameter $x$ is constrained by the following inequalities:
\begin{align}
    x\geq \left(-\frac{(d-2)\mu_CR}{2(d-1)}\right)^{1/d}\equiv x_\text{min}, \qquad -x^d-x^{d+2}-\mu_CRx^2\geq 0. \label{ineq}
\end{align}
As a result, there exist lower and upper bounds for $x$, denoted as $x_\text{min}$ and $x_\text{max}$, respectively.
Here, $x_\text{max}$ corresponds to the value at which the second inequality vanishes.
The value of $x_\text{min}$ indicates the extremal state of dual CFT with maximum free energy, while $x_\text{max}$ represents the thermal state for which $\tilde{\Phi}_2=0$.
Notably, the avoidance of a naked singularity in the bulk is governed by the central charge potential $\mu_C$ (via the constraint $x \geq x_{\text{min}}$) of the dual CFT, rather than the total charge $\tilde{Q}$ as encountered in the conventional RNAdS-BH.
We represent the parameter space for $d=3$, depicting the allowed thermal states of dual CFT (shaded in the light blue region) in terms of $\mu_C$ and $x$ in Fig.~\ref{fig: PS}. 

\begin{figure}[h]
    \centering
    \includegraphics[width = 5 cm]{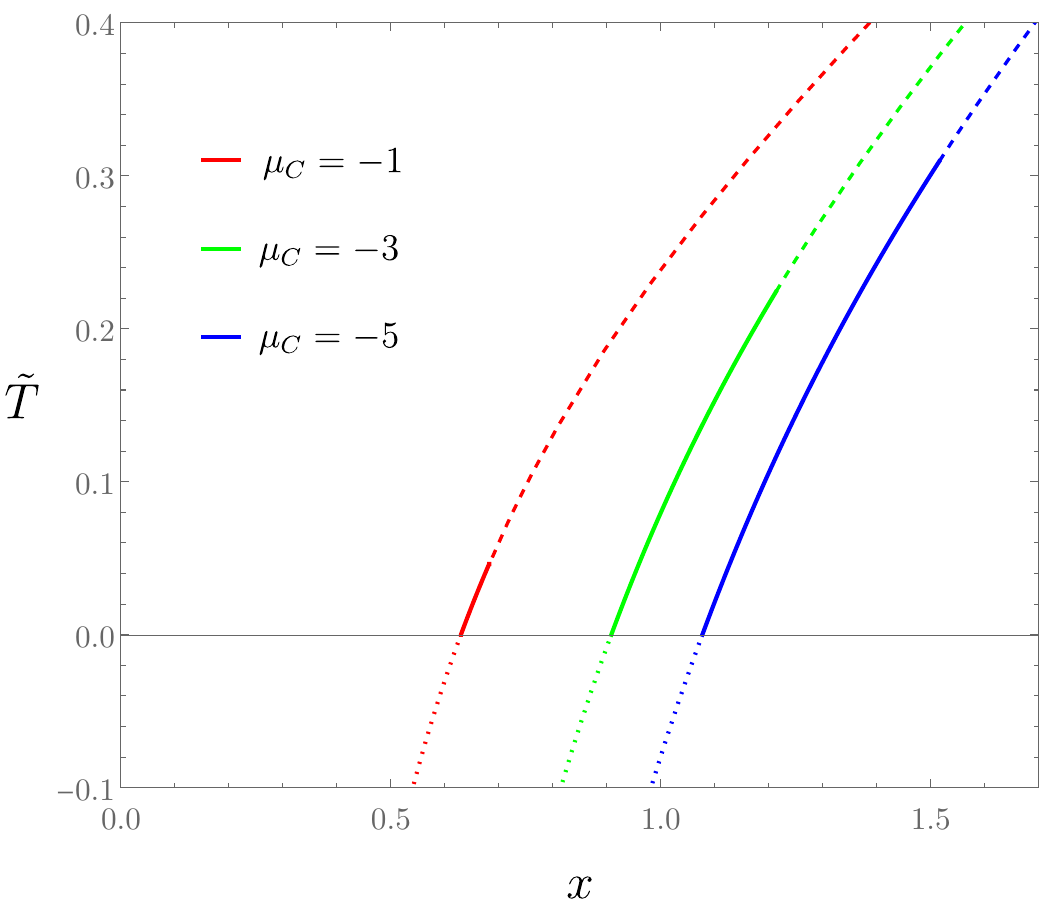} \hspace{0.3cm}
    \includegraphics[width = 5.2 cm]{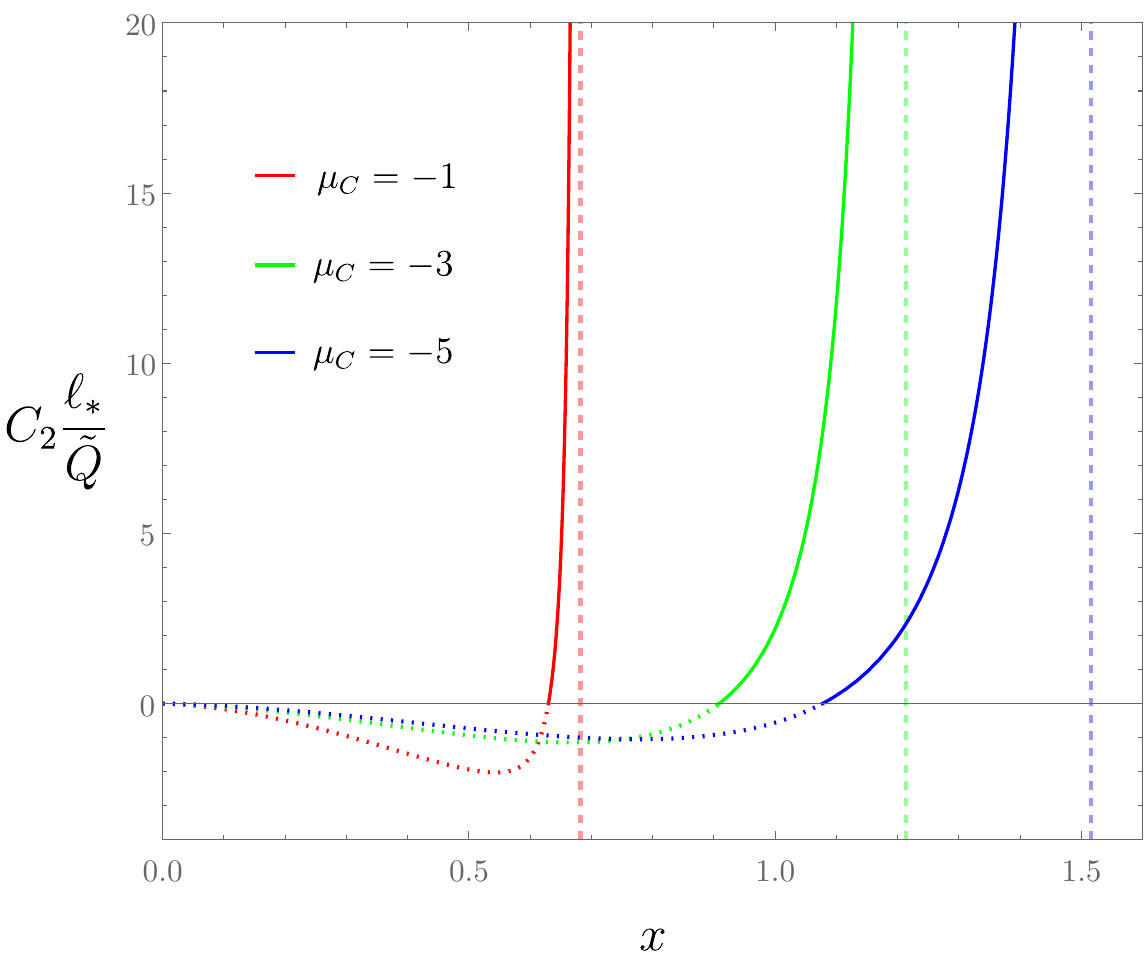}\hspace{0.3 cm}
    \includegraphics[width = 5.2 cm]{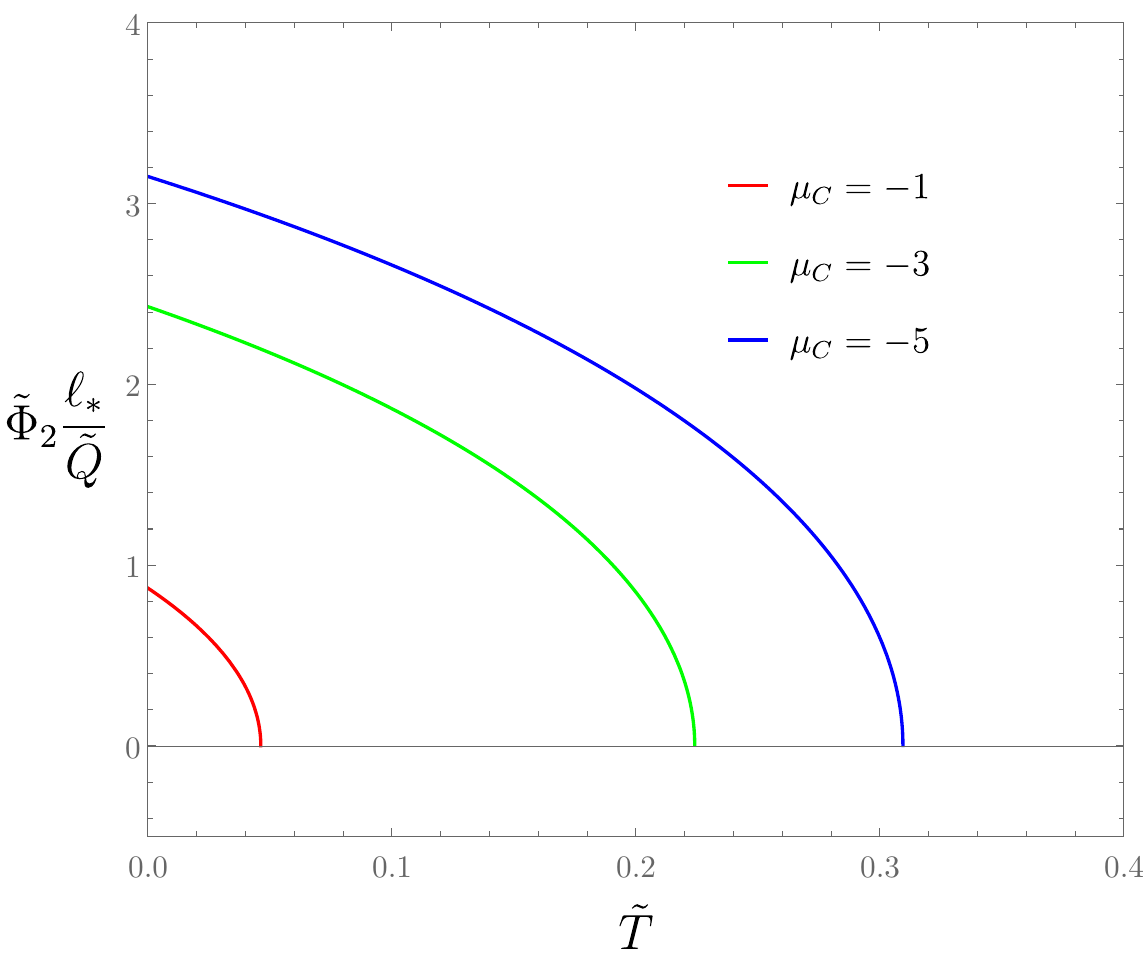}
\caption{The behaviors of temperature $\tilde{T}$ versus $x$ (left panel), the heat capacity $C_2\ell_*/\tilde{Q}$ versus $x$ (middle panel) and Gibbs free energy $\tilde{\Phi}_2\ell_*/\tilde{Q}$ versus temperature $\tilde{T}$ (right panel) of $3$-dimensional dual CFT in the $(\tilde{Q},\mathcal{V},\mu_C)$ ensemble for various values of $\mu_C$ with $R=1$. 
Note the dotted and dashed lines correspond to the regime of $x<x_\text{min}$ and $x>x_\text{max}$, respectively. } 
    \label{fig: 3rd ensemble thermo}
\end{figure}

Fig.~\ref{fig: 3rd ensemble thermo} illustrates the thermodynamic quantities for the 3-dimensional dual CFT with different values of $\mu_C$. 
The left panel illustrates the temperature $\tilde{T}$ as a function of $x$. For a fixed $\mu_C$, $\tilde{T}$ increases with $x$. 
In contrast, when $x$ is kept constant and $\mu_C$ is varied, the results show that $\tilde{T}$ decreases as $\mu_C$ has a more negative value.
Note that $\tilde{T}\leq 0$ when $x\leq x_\text{min}$, and $x_\text{min}$ increases as $\mu_C$ has a more negative value.
The heat capacity $C_2$ versus $x$ is shown in the middle panel.
For $x \leq x_{\text{min}}$, the heat capacity $C_2$ is negative and vanishes at $x = x_{\text{min}}$. 
Within the range $x_{\text{min}} < x < x_{\text{max}}$, the heat capacity becomes positive. At $x = x_{\text{max}}$, $C_2$ diverges and the Gibbs free energy $\tilde{\Phi}_2$ goes to zero, as shown in the right panel. 
Beyond this point, no dual CFT exists.
To compare our study on thermodynamics of the dual CFT in $\mathbb{R} \times \mathbb{H}^{d-1}$ space with that in $\mathbb{R} \times \mathbb{S}^{d-1}$ space \cite{Cong:2021jgb}, the system in hyperbolic space exhibits only a single thermal phase with positive heat capacity. 
As a result, no zeroth-order phase transition exists in this case, unlike the system in the spherical setup.

These findings suggest that the extremal state of the dual CFT corresponds to the extremal BH in the bulk, characterized by vanishing heat capacity.
Notably, the system in the fixed $(\tilde{Q},\mathcal{V},\mu_C)$ ensemble cannot be excited arbitrarily far from its zero temperature state, since an infinite amount of energy would be required to increase the temperature from the state where $\tilde{\Phi}_2= 0$.
In this way, the left and right boundaries of the parameter space in Fig.~\ref{fig: PS} refers to dual CFT with $C_2=0$ $(\tilde{T}=0,\tilde{\Phi}_2=\tilde{\Phi}_\text{max})$ and $C_2=\infty$ $(\tilde{T}=\tilde{T}_\text{max}, \,\tilde{\Phi}_2=0)$, respectively.

\begin{figure}[h]
    \centering
     \includegraphics[width = 5.4 cm]{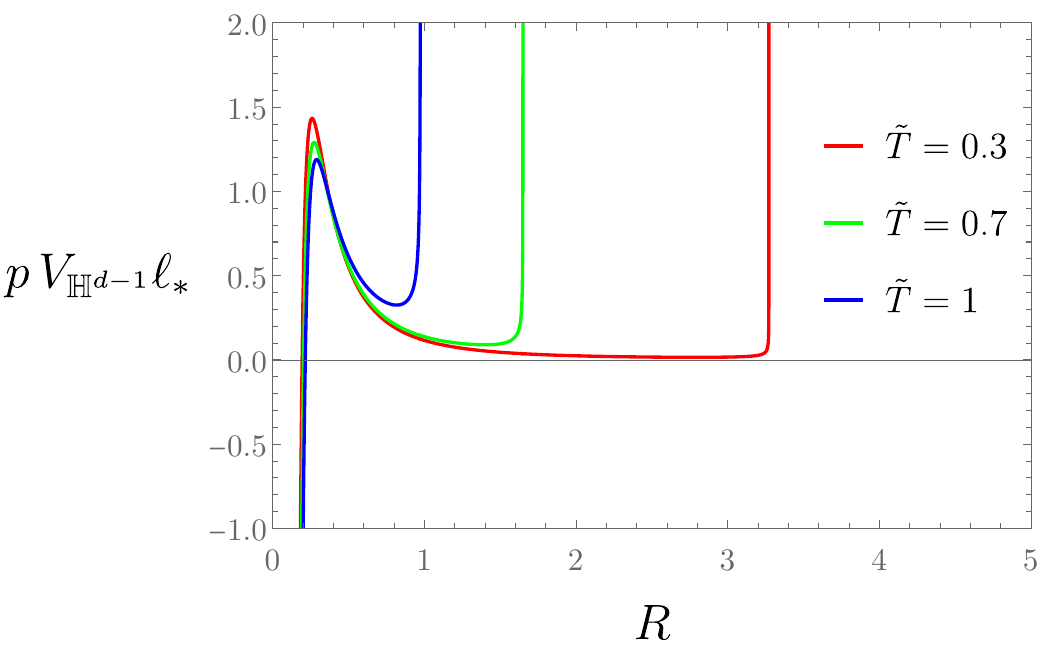}\,\,
     \includegraphics[width = 5.3 cm]{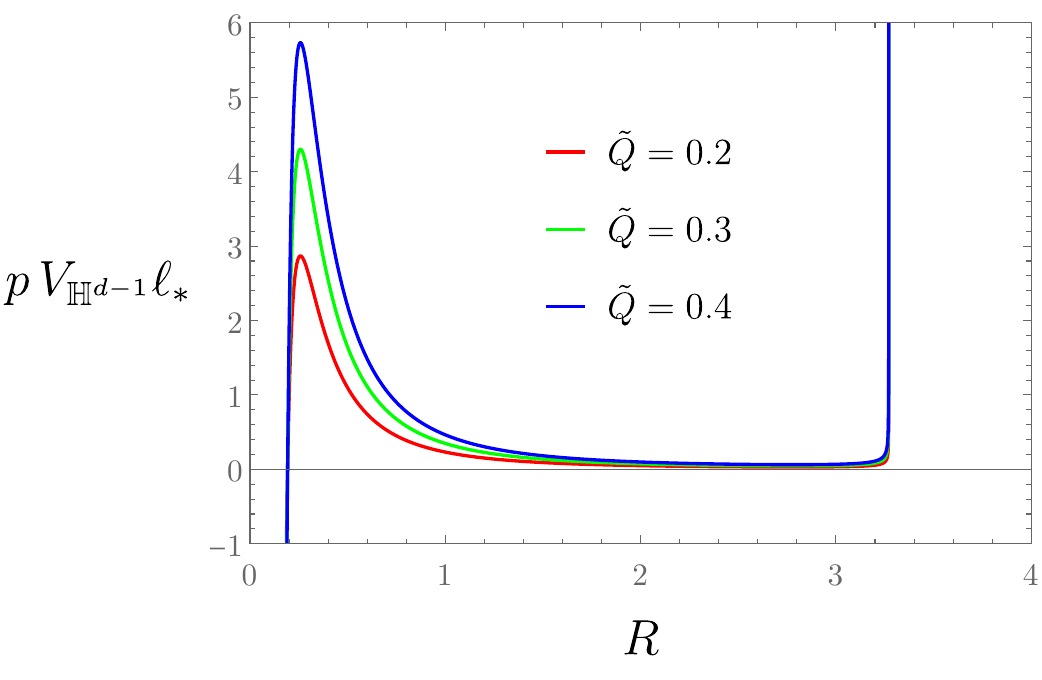}\,\,
     \includegraphics[width = 5.3 cm]{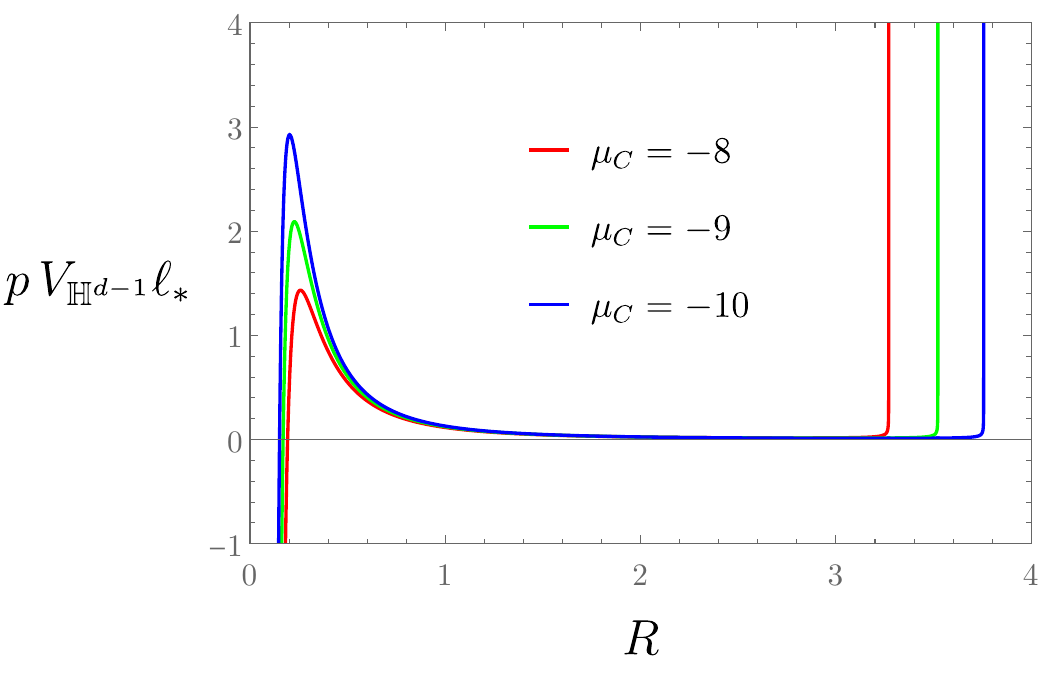}
    \caption{
    The pressure of the 3-dimensional dual CFT versus $x$ for various values of $\tilde{T}$ with fixing $\tilde{Q}=0.1$ and $\mu_C=-8$ (left panel), various values of $\tilde{Q}$ with fixing $\tilde{T}=0.2$ and $\mu_C=-8$ (middle panel) and various values of $\mu_C$ with fixing $\tilde{Q}=0.1$ and $\tilde{T}=0.2$ (right panel).} 
    \label{fig: P-R en2}
\end{figure}

For pressure $p$ in Eq.~\eqref{p charged}, the variables $y$ and $C$ can be eliminated by using Eq.~\eqref{mu C} and the first relation of Eq.~\eqref{charge and potential}. 
As a result, one can write
\begin{eqnarray}
    p=\frac{\tilde{Q}(-2x^d-\mu_CRx^2)}{\VH\ell_*R^dx^{d/2}\sqrt{2(d-1)(d-2)}\sqrt{-x^d-x^{d+2}-\mu_CRx^2}}.
\end{eqnarray}
Using Eq.~\eqref{T3}, the behavior of $p$ as a function of $R$ can be shown in Fig.~\ref{fig: P-R en2}.
Similarly to the previous subsection, the pressure will be negative when $R$ is sufficiently small.
However, it does not increase monotonically in $R$.
As $R$ increases, the pressure in this case will reach its local maximum and then decrease.
Another remarkable point is that the dual CFT cannot be too large.
There is a constraint on $R$ for given $\tilde{T}, \tilde{Q}$ and $\mu_C$.

\newpage

\section{Comment on residual entropy of zero-energy and near-extremal states} \label{sec:Residual Entropy}

In this work,  we adopt a different energy renormalization scheme from~\cite{Emparan:1999gf} by identifying the massless configuration as the ground state (or zero-energy state) of the theory, since it allows the entropy to be expressed via the Cardy-Verlinde formula \cite{Cai:2001jc}. 
Consequently, the extremal configuration, which carries negative energy in this scheme, is identified as the near-extremal state.
However, this identification introduces a subtle tension with the standard Euler relation within the boundary theory.
 
 In this section, we utilize the conformal thermodynamics framework to address this discrepancy. 
 Furthermore, we investigate the mass gap phenomenon in the near-extremal states of the dual CFT within both the fixed $(\tilde{\mu}_Q, \mathcal{V}, C)$ and $(\tilde{Q}, \mathcal{V}, \mu_C)$ ensembles.
 
\subsection{Zero-Energy State}

By introducing $\{\mu_C,C\}$ as a new pair of conjugate variables in the boundary theory, the conformal thermodynamics framework provides a novel explanation of the residual entropy relating to the Euler relation.
A subtlety arises in the zero-energy regime.
The massless configuration of the hyperbolic SAdS-BH possesses a finite entropy even though its vanishing mass. 
From a conventional thermodynamics, this situation appears counterintuitive, since entropy is usually associated with thermal or energetic excitations. 
Moreover, the Euler relation $\tilde{E}= \tilde{T}\tilde{S}_\text{th}$ is no longer satisfied at the point where $\tilde{E}=0$, given that $\tilde{T}$ and $\tilde{S}_\text{th}$ are non-zero.
Within the conformal thermodynamics approach, this behavior admits a satisfactory explanation.
For the zero-energy regime, the Euler relation in Eq.\eqref{p1} yields
\begin{align}
\tilde{E}= \tilde{T}\tilde{S}_\text{th} + \mu_C C=0,
\end{align}
imply that $\tilde{S}_\text{th}$ is reduced to
\begin{align}
\tilde{S}_\text{th}= -\frac{\mu_C C}{\tilde{T}_0}.
\end{align}
As seen in Eq.\eqref{mu C}, $\mu_C$ is non-zero and always negative, so $\tilde{S}_\text{th}> 0$.
This means that the non-zero entropy of the massless bulk configuration arises from the central charge sector rather than $\tilde{E}$.
Hence, the change in entropy is thermodynamically controlled by the first law depending on the process, even when $\tilde{E}$ remains unchanged.

\subsection{Near-Extremal State}

Let us consider some properties for the zero-charge system below the ground state.
When $\tilde{T} \rightarrow 0$, the energy obtained from Eq.\eqref{E CFT} exhibits the power-law behavior of $\tilde{T}$ as follows
\begin{align}
    \tilde{E}=\tilde{E}_\text{ext}+2\pi^2 R^2M_\text{gap}\tilde{T}^2+\mathcal{O}(\tilde{T}^3) \label{extremal Sch}
\end{align}
where
\begin{align}
    \tilde{E}_\text{ext}&=-2\left(\frac{d-1}{d}\right)\left(\frac{d-2}{d}\right)^{\frac{d-2}{2}}\frac{C}{R}, \\
    M_\text{gap}&=|\tilde{E}_\text{ext}|.
\end{align}
Here, $\tilde{E}_\text{ext}$ represents the negative energy of the extremal state of CFT on $\mathbb{R}\times \mathbb{H}^{d-1}$, which holographically corresponds to the mass of the extremal hyperbolic SAdS-BH.
As in the other extremal BHs, we observe the emergence of the \textit{mass gap} $M_{\text{gap}}$, representing a minimum energy required to excite the system from its extremal state, as observed in \cite{Preskill:1991tb, Iliesiu:2020qvm, Emparan:2023ypa}.
The existence of such an energy threshold naturally defines a characteristic temperature scale, denoted by $\tilde{T}_*$. For the neutral case, this is expressed as
\begin{align}
    \tilde{T}_*&=\frac{1}{\sqrt{2}\pi R},
\end{align}
which marks the scale at which thermal excitations become comparable to the $M_{\text{gap}}$.
Therefore, $\tilde{T}_*$ may be interpreted as the threshold temperature required for the emission of a single Hawking quantum $\langle E\rangle \sim k_BT$ from the BH.

Near the extremal state, the corresponding entropy in Eq.\eqref{Sth} and the heat capacity in \eqref{heat capa 1st} can be written in the form
\begin{align}
\tilde{S}_\text{th}
&=\tilde{S}_\text{ext}+4\pi^2R^2M_\text{gap}\tilde{T}+\mathcal{O}(\tilde{T}^2) \label{extremal S Sch} \\
C_1&=4\pi^2R^2M_\text{gap}\tilde{T}+\mathcal{O}(\tilde{T}^2), \label{extremal heat capa Sch}
\end{align}
where the residual entropy reads
\begin{align}
    \tilde{S}_\text{ext}= 4\pi C 
\left(\frac{d-2}{d}\right)^{\frac{d-1}{2}}.
\end{align}
The expression above confirms that the entropy approaches the finite value $\tilde{S}_\text{ext}$, while the heat capacity vanishes linearly as $\tilde{T} \rightarrow 0$.
This behavior does not contradict the Nernst form of the third law since it constrains the vanishing of heat capacity rather than the entropy itself. 
Instead, the finite residual entropy reflects a microscopic degeneracy at absolute zero, which can be found in many systems~\cite{PhysRev.162.162, Lau_2006, Morita_2016, Ohkawa_2012}.

The existence of a negative energy $\tilde{E}_\text{ext} < 0$ suggests an inadequacy in the conventional form of the Euler relation in dual CFTs.
Namely, the term $\tilde{T}\tilde{S}_\text{th}$ vanishes in the zero-temperature limit, it fails to account for the non-vanishing negative value of $\tilde{E}_\text{ext}$.
Within the framework of conformal thermodynamics, the Euler relation is generalized to include the central charge sector as 
\begin{align}
    \tilde{E}_\text{ext}=\mu_C C<0,
\end{align}
which implies a negative central charge potential $\mu_C<0$.
In this way, the Euler relation remains consistent even in the extremal limit.
Consequently, energy variations in this regime are driven by the central charge sector rather than thermal excitations, allowing the entropy to remain constant while the heat capacity approaches zero.
This perspective provides a complementary thermodynamic interpretation to the analysis of hyperbolic BH in~\cite{Emparan:1999gf}, where the residual entropy is primarily discussed in terms of geometric and entanglement properties.

In preparation for our analysis of the thermal crossover in holographic R\'enyi entropy, it is essential to characterize the low-temperature behavior of the dual charged CFT on $\mathbb{R} \times \mathbb{H}^{d-1}$. 
To this end, we provide a detailed analysis within both thermodynamic ensembles.
For the fixed $(\tilde{\mu}_Q,\mathcal{V},C)$ ensemble, we substitute $x$ from Eq.\eqref{x for 1st en} into the expression for the internal energy $\tilde{E}_1$, thermal entropy $\tilde{S}_\text{th}$ and heat capacity $C_1$. 
By expanding them in a low-temperature limit, we find that these thermodynamic quantities take the forms given in Eqs.\eqref{extremal Sch}, \eqref{extremal S Sch} and \eqref{extremal heat capa Sch}, respectively, with the following parameters 
\begin{align}
    \tilde{E}_\text{ext}&=-2\left(\frac{d-1}{d}\right)\left(\frac{d-2}{d}\right)^{\frac{d-2}{2}}\frac{C}{R}\Delta^{d/2},     \\
    \tilde{S}_\text{ext}&=4\pi C\left[ \frac{(d-2)\Delta}{d}\right]^{\frac{d-1}{2}}. \\
    M_\text{gap}&=\frac{|\tilde{E}_\text{ext}|}{\Delta}=2\left(\frac{d-1}{d}\right)\left(\frac{d-2}{d}\right)^{\frac{d-2}{2}}\frac{C}{R}\Delta^{\frac{d-2}{2}}, \\
    \tilde{T}_*&=\frac{\sqrt{\Delta}}{\sqrt{2}\pi R}, \label{T* 1st en}
\end{align}
Here, $\Delta$ defined in Eq.\eqref{def delta} accounts for the effects of $\tilde{\mu}_Q$ on the extremal configuration.
As $\tilde{\mu}_Q$ increases, the higher value of $M_\text{gap}$ signifies a more robust vacuum configuration, while the enhancement of $\tilde{S}_\text{ext}$ suggests a higher degeneracy at the extremity.

Next, we consider a novel fixed $(\tilde{Q}, \mathcal{V}, \mu_C)$ ensemble for the charged CFT in the near-extremal limit. 
From the expression of temperature in Eq.\eqref{T3}, we defined $h(x)$ such that
\begin{align}
    4\pi R\,\tilde{T}=h(x)=2(d-1)x+\frac{(d-2)\mu_CR}{x^{d-1}}.
\end{align}
Since $h(x_{\text{min}})=0$ at $x= x_{\text{min}}$, we perform a Taylor expansion of $h(x)$ around $x_{\text{min}}$ as follows
\begin{align}
    4\pi R\, \tilde{T}=h(x_\text{min})+h'(x_\text{min})(x-x_\text{min})+\cdots
\end{align}
This yields the linear relationship between the parameter $x$ and the temperature $\tilde{T}$
\begin{align}
    x(\tilde{T})\approx x_\text{min}+\frac{4\pi R}{h'(x_\text{min})}\tilde{T}=x_\text{min}+\frac{2\pi R}{d(d-1)}\tilde{T}. \label{x expand}
\end{align}
By substituting the expansion of $x(\tilde{T})$ into the expression for the internal energy $\tilde{E}_2$ as expressed in Eq.\eqref{E2}, we find that it takes a form similar to the two previous cases in the low-temperature limit.
Consequently, the extremal energy and its corresponding residual entropy are given by
\begin{align}
    \tilde{E}_\text{ext}&=\left(\frac{d-1}{d-2}\right)^{1/2}\left(\frac{d}{d-2}x_\text{min}^2-1\right)^{1/2}\frac{\sqrt{2}\tilde{Q}}{R\ell_*}, \label{extremal 2 ensemble} \\
    \tilde{S}_\text{ext}&=\frac{4\pi \tilde{Q}^2x_\text{min}}{(d-2)\ell_*^2R\tilde{E}_\text{ext}},
\end{align}
respectively.
It is important to note that this ensemble is permitted to investigate the behavior of the central charge $C$ as a function of temperature.
Substituting $x(\tilde{T})$ in Eq.\eqref{x expand} into the expression for $C$ in Eq.\eqref{C ensemble},
we have
\begin{align}
    C=C_\text{ext}+\frac{4\pi\tilde{Q}^4}{d(d-2)\ell_*^4R^2\tilde{E}_\text{ext}^3x_\text{min}^{d-1}}\,\tilde{T}+\mathcal{O}(\tilde{T}^2), \label{C_expansion}
\end{align}
where the residual central charge is given by
\begin{align}
    C_\text{ext}=\frac{\tilde{Q}^2}{(d-2)\ell_*^2R\tilde{E}_\text{ext}x_\text{min}^{d-2}}.
\end{align}
Three novel physical implications arise from these results. 
First, we observe that the extremal energy $\tilde{E}_{\text{ext}}$ is positive throughout the allowed parameter space. 
It contrasts to dual CFT on hyperbolic space, where the extremal energy is typically negative.
However, the system exhibits a mass gap, as evidenced by the energy excitation scaling $\Delta \tilde{E}=\tilde{E}-\tilde{E}_\text{ext}\sim \tilde{T}^2$. 
Identifying $2\pi^2R^2M_\text{gap}$ as the coefficient of $\tilde{T}^2$, we obtain
\begin{align}
    M_\text{gap}&=\frac{\sqrt{2}\tilde{Q}}{R\ell_*} \frac{\left[d(d-1)x_\text{min}^2 - (d-2)\right]}{d\left[(d-1)(dx_\text{min}^2 - (d-2))\right]^{3/2}}. \label{mgap 2nd}
\end{align}
Since the internal energy scales extensively with the charge $\tilde{Q}$, the typical energy associated with a Hawking quantum is naturally characterized by the scale $\langle E \rangle \sim \tilde{Q}\tilde{T}$. 
The characteristic temperature $\tilde{T}_*$ can be obtained by equating the thermal energy scale to the macroscopic excitation energy, which yielding
\begin{align}
    \tilde{T}_* &=\frac{\tilde{Q}}{2\pi^2 R^2 M_\text{gap}}=\frac{\ell_*}{2^{3/2}\pi^2 R} \frac{d\left[(d-1)(dx_\text{min}^2 - (d-2))\right]^{3/2}}{\left[d(d-1)x_\text{min}^2 - (d-2)\right]}.\label{T* 2nd en}
\end{align}

Second, a near-extremal structure of the charged CFT was modified from a neutral case by a central charge sector rather than a charge sector as encountered in conventional CFT, which is holographically dual to RNAdS-BH.
In particular, from Eqs.\eqref{extremal 2 ensemble}-\eqref{T* 2nd en}, a possible value of these quantities was restricted by the value of $\mu_C$ as follows
\begin{align}
    |\mu_C|R> 2\left(\frac{d-1}{d}\right)\left(\frac{d-2}{d}\right)^{\frac{d-2}{2}}.
\end{align}

Third, the expansion in Eq.\eqref{C_expansion} indicates that while $C$ reaches a non-zero value $C_{\text{ext}}$ at the extremal limit, it receives a leading-order thermal correction that is linear in $\tilde{T}$. 
This suggests that the degrees of freedom in the dual CFT are thermally excited as the system departs from the absolute zero temperature.
\begin{figure}[h]
\centering
\includegraphics[width = 6 cm]{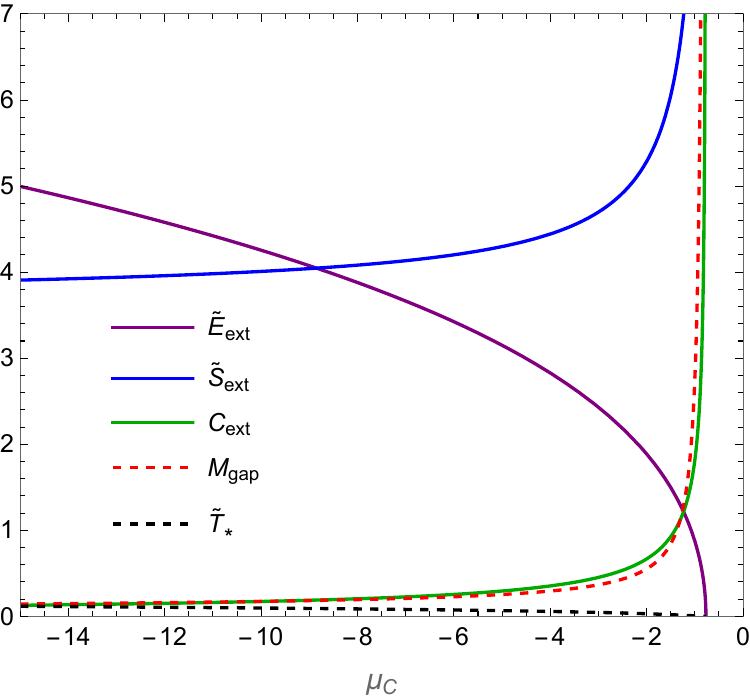}
\caption{Plot $\tilde{E}_\text{ext}$, $S_\text{ext}$, $C_\text{ext}$, $M_\text{ext}$ and $\tilde{T}_*$ against central $\mu_C$ for dual CFT in the ensemble of fixed $(\tilde{Q},\mathcal{V},\mu_C)$. Note that we use $\tilde{Q}=1,R=1$ and $\ell_*=1$ in these plot.} 
\label{fig: extremal}
\end{figure}

Figure~\ref{fig: extremal} illustrates the dependence of the residual thermodynamic quantities (solid curves) and characteristic energy scales (dashed curves) on the central charge potential $\mu_C$ in a statistical ensemble of holographic CFTs with different numbers of degrees of freedom. 
The monotonic increase of the residual central charge $C_{\rm ext}$ with $\mu_C$ ensures the thermodynamic stability condition $\partial C/\partial\mu_C>0$, analogous to the positivity of $\partial N/\partial\mu>0$ in conventional thermodynamics.

\newpage
\section{Characteristic R\'enyi Index} \label{sec:index}
\subsection{Holographic Calculation of R\'enyi Entropy}
Consider a quantum field theory in the vacuum state of a $d$-dimensional Minkowski spacetime.
The system is divided into two subsystems, $A$ and $B$, by a $(d-2)$-dimensional hypersurface $\Sigma$, known as the entangling surface. 
The $n$-th R\'enyi entropy across $\Sigma$ is defined as \cite{renyi1961measures}
\begin{eqnarray}
S_n=\frac{1}{1-n}\log \text{Tr}\left(\rho_A^n\right), \label{Renyi}
\end{eqnarray}
where $n$ is the R\'enyi index and $\rho_A=\text{Tr}_B\rho$ is the reduced density matrix of subsystem $A$.
Some specific values of $n$ capture distinct the information about the $\rho_A$.
The entanglement entropy (EE) is obtained from the limit $n\to1$,
\begin{eqnarray}
S_{\rm EE}=\lim_{n\rightarrow 1}S_n .
\end{eqnarray}
The limit $n\rightarrow \infty$ yields the Min-entropy
\begin{align}
    S_\infty =\lim_{n\rightarrow \infty}S_n=-\ln \lambda_\text{max}
\end{align}
where $\lambda_\text{max}$ is the largest eigenvalue of $\rho_A$.
The limit $n\rightarrow 0$ gives the Hartley entropy
\begin{align}
    S_0=\lim_{n\rightarrow 0}S_n=\ln D
\end{align}
where $D$ is the number of non-vanishing eigenvalues of $\rho_A$.

In general, calculating $\rho_A$ by performing a partial trace over the states outside region A is a non-trivial task. To bypass this direct calculation, one can recast the problem into a thermodynamics-like description by introducing  the modular Hamiltonian $K_A$ as follows
\begin{align}
    \rho_A=\frac{e^{-K_A}}{\text{Tr}(e^{-K_A})}. \label{eq:Hmod}
\end{align}
While $K_A$ is typically a non-local and complicated operator for generic quantum systems, an elegant simplification occurs for the vacuum state of CFT when region $A$ is a spherical ball of radius $R$ in Minkowski spacetime. 
Through the CHM map, the causal diamond of this spherical region can be conformally mapped to a hyperbolic cylinder $\mathbb{R} \times \mathbb{H}^{d-1}$. 
Consequently, the EE across the region $A$ in the Minkowski vacuum is mapped to the thermal entropy of the same CFT on the hyperbolic cylinder $\mathbb{R}\times \mathbb{H}^{d-1}$ at temperature $T_0=1/(2\pi R)$. 
This relation is geometrically encoded in the fact that the modular flow is mapped to time translations on $\mathbb{R}\times \mathbb{H}^{d-1}$. 
Specifically, the unit hyperbolic metric defined in Eq.~\eqref{unit hyp metric1} can be written as
\begin{eqnarray}
ds_{\mathbb{H}^{d-1}}^2=\frac{dz^2}{z^2-1}+(z^2-1)ds^2_{\mathbb{S}^{d-2}} .
\end{eqnarray}
Here the coordinate $z$ is related to the radial coordinate $u$ through $z=\cosh u$, and the volume of the unit $(d-2)$-sphere is 
\begin{eqnarray}
V_{\mathbb{S}^{d-2}}=\frac{2\pi^{(d-1)/2}}{\Gamma \big[(d-1)/2\big]} .
\end{eqnarray}
Since $\mathbb{H}^{d-1}$ has infinite volume, one introduces a cutoff to regulate it,
\begin{eqnarray}
V_{\mathbb{H}^{d-1}} =V_{\mathbb{S}^{d-2}}\int_1^{z_\text{max}}dz\,(z^2-1)^{\frac{d-3}{2}},
\end{eqnarray}
where $z_\text{max}=R/\delta$.
In the holographic framework, the thermal entropy of the dual conformal field theory can be computed from the area of the black hole horizon in the bulk. 
For the case corresponding to the CHM map, the relevant bulk geometry is a massless hyperbolic black hole ($m=0$). 
The resulting thermal entropy is
\begin{eqnarray}
S_\text{th}=\frac{L^{d-1}}{4G_\text{N}}V_{\mathbb{H}^{d-1}} .
\end{eqnarray}
For $d=2$, this expression reproduces the well-known entanglement entropy of a single interval \cite{Calabrese:2009qy,Calabrese:2004eu}
\begin{eqnarray}
S_\text{EE} =\frac{c}{3}\log \left( \frac{2R}{\delta }\right),
\end{eqnarray}
where the central charge $c=3L/2G_\text{N}$ follows from the Brown–Henneaux relation in $\text{AdS}_3/\text{CFT}_2$ \cite{Brown:1986nw}.

To compute the $n$-th R\'enyi entropy, the CHM map unitarily transforms the reduced density matrix into a thermal density matrix, leading to
\begin{eqnarray}
\text{Tr}\left(\rho_A^n\right)=\frac{Z(T_0/n)}{Z(T_0)^n},
\end{eqnarray}
where $Z(T_0)$ is the thermal partition function at temperature $T_0=1/(2\pi R)$. 
The partition function $Z(T_0/n)$ therefore corresponds to the same theory evaluated at the rescaled temperature $T_0/n$.
Substituting the above relation into Eq.~\eqref{Renyi}, we obtain 
\begin{eqnarray}
    S_n&=&\frac{1}{1-n}\log \frac{Z(T_0/n)}{Z(T_0)^n}
    =\frac{1}{1-n}\left[ \log Z(T_0/n)-n\log Z(T_0)\right]. \nonumber
\end{eqnarray}
From conventional thermodynamics, the free energy can be obtained from the partition function as $F=-T\log Z(T)$.
The above relation can be written in the form \cite{Baez:2011upp}
\begin{eqnarray}
    S_n=-\frac{F(T_0)-F\left(T_0/n\right)}{T_0-T_0/n}
    =\frac{1}{T_0-T_0/n}\int_{T_0/n}^{T_0}S_\text{th}(T)dT. \label{Renyi 2}
\end{eqnarray}
Note that the last equality has been obtained by using the relation between the thermal entropy $S_\text{th}$ and the free energy $F$ in the form $S_\text{th}=-\partial F/\partial T$.

The calculation of the free energy of a CFT on hyperbolic space remains difficult to compute directly.
In the AdS/CFT correspondence, the partition function of the boundary CFT is equivalent to the gravitational partition function in the bulk spacetime. 
Thus, the free energy of the CFT can be obtained from the on-shell action of the corresponding bulk solution. 
In the present case, the relevant bulk geometry is a hyperbolic black hole. 
Therefore, the R\'enyi entropy can be evaluated holographically by substituting $F(T)$ in Eq.\eqref{Renyi 2} with the free energies of a hyperbolic black hole at two different temperatures, $T_0=1/(2\pi R)$ and $T_0/n$, where  $T_0$ is the Hawking temperature of pure AdS spacetime written in hyperbolic slicing.

Consequently, the holographic R\'enyi entropy inherits fundamental property constraints from both quantum information theory and bulk thermodynamics. To ensure the physical viability of the density matrix eigenvalues and the thermodynamic stability of the dual hyperbolic black hole, the R\'enyi entropy $S_n$ must satisfy a set of general differential inequalities for all $n > 0$. These constraints, known as the R\'enyi entropy inequalities, are explicitly given by
\begin{align}
    \frac{\partial S_n}{\partial n}&\leq 0 && \text{Inequality I}, \label{eq:I}\\
    \frac{\partial}{\partial n}\left(\frac{n-1}{n}S_n\right)&\geq 0 && \text{Inequality II}, \\ 
    \frac{\partial}{\partial n}\big((n-1)S_n\big)&\geq 0 && \text{Inequality III}, \\
    \frac{\partial^2}{\partial n^2}\big((n-1)S_n\big)&\leq 0 && \text{Inequality IV}.\label{eq:IV}
\end{align}
These inequalities carry profound physical interpretations linking boundary quantum information to bulk thermodynamics. Inequality I dictates that $S_n$ monotonically decreases with the index $n$, reflecting the algebraic ordering of the density matrix spectrum. In the thermal description, Inequalities II and III establish the fundamental positivity of the thermal entropy and the thermal energy, respectively. Finally, the concavity condition of Inequality IV corresponds to the positivity of the modular energy variance (the capacity of entanglement). This directly translates to a positive heat capacity, which rigorously guarantees the local thermodynamic stability of the dual hyperbolic black hole against thermal fluctuations.

\newpage

\subsection{The Crossover in Charged R\'enyi Entropy}\label{sec:Renyi}

Having established $\tilde{T}_*$ as the characteristic scale associated with the mass gap, we now investigate how this scale is encoded in the entanglement structure of a spherical reduced state through the holographic R\'enyi entropy.
In particular, we introduce a physically motivated partition of the spectrum, defined by a characteristic R\'enyi index $n_*$ associated with $\tilde{T}_*$. 
This partition separates a sector predominantly controlled by the largest eigenvalues from a regime where subleading eigenvalues contribute more significantly.
We interpret the R\'enyi entropy of order $n$ as a probe of different sectors of the entanglement spectrum of the reduced density matrix $\rho_A$.
For $n>n_*$, the R\'enyi entropy predominantly probes the sector associated with the largest eigenvalues of $\rho_A$, corresponding to low modular-energy states that capture the vacuum-like structure of subsystem $A$.
In contrast, for $n<n_*$, the R\'enyi entropy becomes increasingly sensitive to smaller eigenvalues, where thermal-like modular excitations contribute significantly.
Through the CHM map, this construction admits a natural interpretation as encoding the crossover between near-extremal and thermally excited regimes in the dual hyperbolic CFT at $\tilde{T} \sim \tilde{T}_*$.

In this section, we first study how this partition affects charge R\'enyi entropy in the fixed $(\tilde{\mu}_Q,\mathcal{V},C)$ ensemble. 
We then turn to the investigation of the holographic R\'enyi entropy in the novel ensemble with fixed $(\tilde{Q},\mathcal{V},\mu_C)$ in the thermal CFT.


Using the expression in Eq.~\eqref{Renyi 2}, one can obtained $S_n$ via the thermodynamic quantities (with tildes) of dual CFT as follows
\begin{align}
    S_n(\tilde{\mu}_Q)&=\frac{1}{\tilde{T}_0-\tilde{T}_0/n}\int_{\tilde{T}_0/n}^{\tilde{T}_0}\tilde{S}_\text{th}(\tilde{T})d\tilde{T} \nonumber \\
    &=\frac{n}{n-1}\frac{1}{\tilde{T}_0}\int_{x_n}^{x_1}\tilde{S}_\text{th}\frac{d\tilde{T}}{dx}dx \nonumber \\
    &=\frac{n}{n-1}\frac{S_\text{EE}}{2}\left[ (x_1^{d-2}-x_n^{d-2})\Delta +x_1^d-x_n^d\right]
    \label{Sn_1st}
\end{align}
Here, $x_n$ is the largest solution of $\tilde{T}(x_n,\tilde{\mu}_Q)=\tilde{T}_0/n$, which yields
\begin{align}
    x_n=\frac{1}{dn}+\sqrt{\frac{1}{d^2n^2}+\frac{d-2}{d}\Delta}.
\end{align}
We refer $S_\text{EE}$ to the entanglement entropy such that $S_\text{EE}=S_1(\tilde{\mu}_Q=0)$.
The behaviors of $S_n(\tilde{\mu}_Q)$ in terms of $n$ and $\tilde{\mu}_Q$ are explored in \cite{Belin:2013uta}.

As discussed in Section~\ref{sec:Residual Entropy}, the characteristic temperature $\tilde{T}_{*}$ marks the scale at which the system transitions from a near-extremal regime to a regime where thermal excitations become significant. 
In the holographic calculation of R\'enyi entropy, this temperature scale determines the R\'enyi index $n_*$ through the relation  
\begin{align}
    n_*=\frac{\tilde{T}_0}{\tilde{T}_*}=\frac{1}{\sqrt{2\Delta}}, \label{eq:n*}
\end{align}
Note that $\tilde{T}_*=\tilde{T}_0/n_*$.
In the context of the reduced density matrix $\rho_A$ for a spherical region in the Minkowski vacuum, we interpret $n_*$ as the threshold separating two distinct regimes within the entanglement spectrum.
Specifically, the R\'enyi entropy of order $n>n_*$ probes the sector associated with the largest eigenvalues of $\rho_A$. 
These large eigenvalues correspond to low-energy eigenstates of the modular Hamiltonian, thereby capturing the modular vacuum structure of subsystem $A$.
In contrast, the R\'enyi entropy of order $n<n_*$ becomes increasingly sensitive to the smaller eigenvalues of $\rho_A$, which correspond to higher-energy modular eigenstates where the contribution of thermal-like modular excitations becomes important.
Therefore, $n_*$ effectively characterizes a crossover between a regime dominated by the vacuum-like sector of the entanglement spectrum and a regime where thermal-like excitations play a significant role. 
In the thermodynamic picture, these two regimes map to the near-extremal and thermally excited sectors of the bulk black hole.
Note that the thermal-like excitations appearing at $n < n_*$ reflect the emergence of Hawking radiation as the black hole moves away from the extremal limit.

For $\tilde{\mu}_Q=0$, Eq.~\eqref{eq:n*} gives $n_*=\sqrt{2}/2 \approx 0.707$, which is independent of the dimension $d$ of the system. 
From the expression of R\'enyi entropy in Eq.~\eqref{Sn_1st}, we observe that the R\'enyi entropy of order $n_*$ normalized by the entanglement entropy ($S_{n_*}(0)/S_\text{EE}$) yields $1.207, 1.229, 1.234$, and $1.236$ for $d=2, 3, 4$, and $5$, respectively.
In the limit of $d\rightarrow \infty$, we have
\begin{align}
    \frac{S_{n_*}(0)}{S_\text{EE}}=\frac{\sqrt{2} \left(e^{\sqrt{2}-1}-1\right)}{2-\sqrt{2}}=1.239
\end{align}

\begin{figure}[H]
\centering
\includegraphics[width = 5.6 cm]{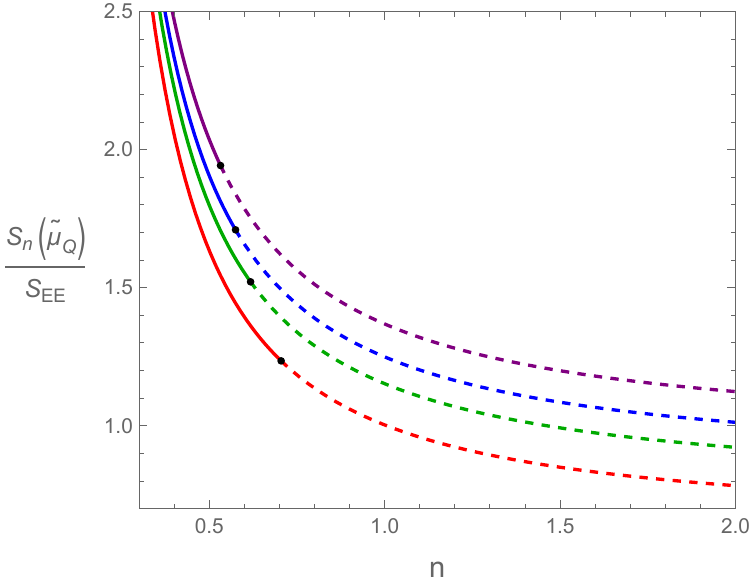}\hspace{1cm}
\includegraphics[width = 5.5 cm]{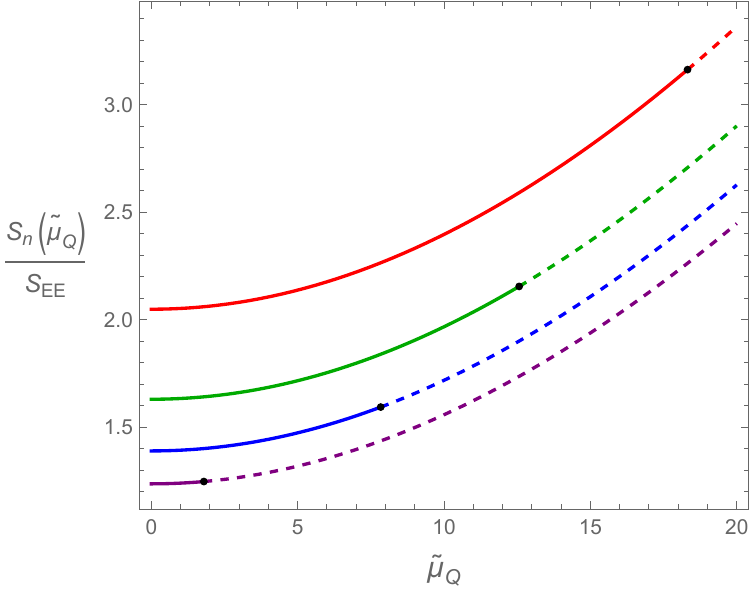}
\caption{\textbf{Left:} Behavior of the normalized R\'enyi entropy $S_n(\tilde{\mu}_Q)/S_\text{EE}$ as a function of $n$ for $d=3$ dimensional system. 
The red, green, blue, and purple curves are correspond to fixed $\tilde{\mu}_Q = 1, 7, 9,$ and $11$ respectively.
\textbf{Right:} Plots of $S_{n}(\tilde{\mu}_Q)/S_\text{EE}$ as a function of $\tilde{\mu}_Q$ for fixed $n=0.4$ (red) $0.5$ (green) $0.6$ (blue) and $0.7$ (purple).
}
\label{fig: cross 1st}
\end{figure}

\begin{figure}[H]
\centering
\includegraphics[width = 5.2 cm]{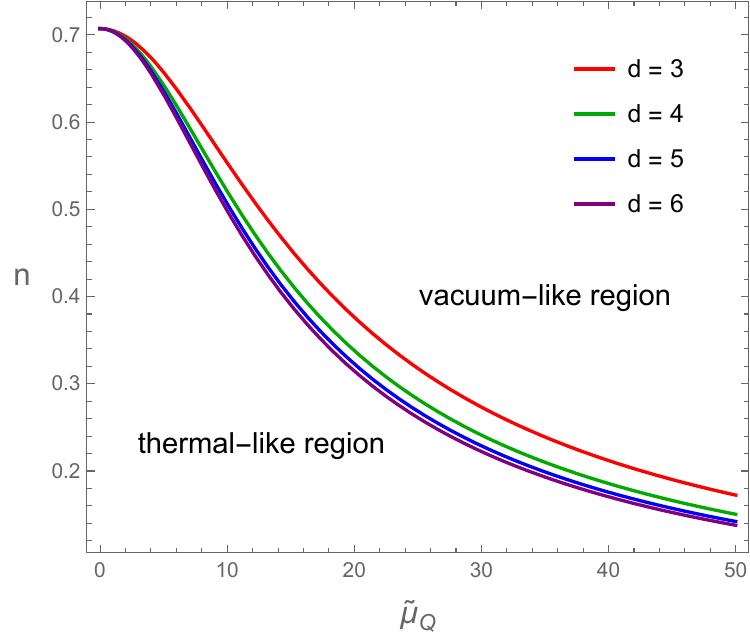}\hspace{1cm}
\includegraphics[width = 5.8 cm]{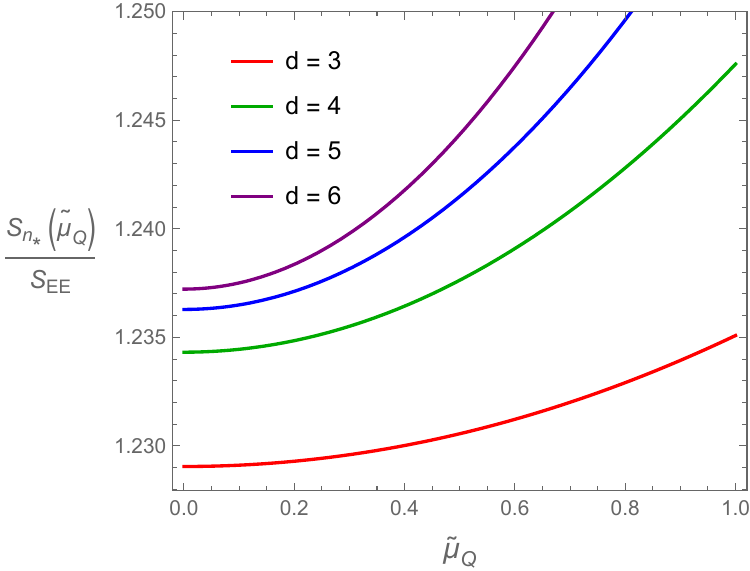}
\caption{\textbf{Left:} The characteristic index $n_*$ is plotted in the R\'enyi index space of charge R\'enyi entropy for $d=3$ (red), $4$ (green), $5$ (blue) and $6$ (purple).
\textbf{Right:} Plots of $S_n(\tilde{\mu}_Q)/S_\text{EE}$ at $n=n_*$ versus $\tilde{\mu}_Q$ for $d=3$, $4$, $5$ and $6$ in the same color code.} 
\label{fig:SnvsmuQ}
\end{figure}

We illustrate the ratio $S_n(\tilde{\mu}_Q)/S_\text{EE}$ versus the R\'enyi index $n$ with constant chemical potential $\tilde{\mu}_Q$ in the left panel in Fig.~\ref{fig: cross 1st}.
The red, green, blue, and purple curves represent $\tilde{\mu}_Q = 1, 7, 9,$ and $11$, respectively. 
The characteristic scale $n_*$ (marked by black dots) divides the curves into two distinct sectors, i.e., $n < n_*$ (solid curves) and $n > n_*$ (dashed curves). 

The results for the R\'enyi entropy of order $n$ as a function of $\tilde{\mu}_Q$ are displayed in the right panel in Fig.~\ref{fig: cross 1st}.
The red, green, blue and purple curves are correspond to $n=0.4$, $0.5$, $0.6$ and $0.7$, respectively.
For varying $\tilde{\mu}_Q$ at fixed R\'enyi index $n$, the R\'enyi entropy can probe two distinct sectors of the entanglement spectrum of $\rho_A$, depending on the value of the chemical potential. Specifically, the R\'enyi entropy of order $n$ predominantly probes the thermal-like sector when $\tilde{\mu}_Q<\tilde{\mu}_{Q*}$ (solid curves), while it probes the vacuum-like sector for $\tilde{\mu}_Q>\tilde{\mu}_{Q*}$ (dashed curves).
The chemical potential $\tilde{\mu}_{Q*}$ (marked by black dots) can be obtained from Eq.~\eqref{eq:n*} by imposing $n_*=n$ and $\tilde{\mu}_Q=\tilde{\mu}_{Q*}$, yielding
\begin{align}
    \tilde{\mu}_{Q*}=\frac{2\pi}{R\ell_*}\sqrt{\frac{(d-1)(1-2n^2)}{(d-2)n^2}}. \label{eq:muQstar}
\end{align}

Here, we introduce the term \textit{R\'enyi index space} to denote the parameter space spanned by the R\'enyi index $n$ and its relevant thermodynamic parameters, such as the $n$--$\tilde{\mu}_Q$ plane considered here. 
In this space, the $n_*$ curve separates regions in which the R\'enyi entropy probes different sectors of the entanglement spectrum of the reduced density matrix $\rho_A$.
In particular, R\'enyi entropies with $(n,\tilde{\mu}_Q)$ located above the $n_*$ curve are predominantly sensitive to the largest eigenvalues of $\rho_A$, corresponding to the vacuum-like sector of the modular spectrum. 
In contrast, points lying below the $n_*$ curve receive increasingly significant contributions from smaller eigenvalues, where thermal-like modular excitations become important.

In the left panel of Fig.~\ref{fig:SnvsmuQ}, we show the characteristic index $n_*$ curves in the $n$--$\tilde{\mu}_Q$ plane for $d=3,4,5,$ and $6$, corresponding to the red, green, blue, and purple curves, respectively. 
As $\tilde{\mu}_Q$ increases, the vacuum-like sector of the modular spectrum is observed to expand in the R\'enyi index space.
This behavior follows from the monotonic decrease of the characteristic R\'enyi index $n_*$ with increasing $\tilde{\mu}_Q$. 
Notably, from Eq.\eqref{eq:muQstar}, the real-valued $\tilde{\mu}_{Q*}$ exists only for $n<1/\sqrt{2}\approx 0.707$.
This means that R\'enyi entropy of order $n\geq 1/\sqrt{2}$ entirely probes the vacuum-like sector of subsystem $A$ regardless of its chemical potential.
Furthermore, the value of the R\'enyi entropy evaluated at $n_*$ increases with $\tilde{\mu}_Q$, as shown in the right panel of Fig.~\ref{fig:SnvsmuQ}.

Motivated by these holographic results, it is natural to ask how a finite-order R\'enyi entropy can probe the large-eigenvalue sector of the entanglement spectrum, a regime conventionally associated with the limit $n\rightarrow \infty$. 
Through the CHM map, the hyperbolic thermal description can be mapped back to a spherical subregion in the Minkowski vacuum, thus clarifying the physical role of $\tilde{\mu}_Q$ in the original field theory setup.
Since $\rho_A$ is generated by the modular Hamiltonian $K_A$ as defined in Eq.\eqref{eq:Hmod}, introducing  $\tilde{\mu}_Q$ amounts to generalizing $K_A$ into a grand canonical form
\begin{align}
    K_{\text{gen}} &= K_A - \frac{\tilde{\mu}_Q}{\tilde{T}_0} \tilde{Q}_A,
\end{align}
where $\tilde{Q}_A$ is the local charge operator enclosed within the spherical region, and $\tilde{T}_0$ acts as the modular temperature. 
The corresponding density matrix $\rho_A(\tilde{\mu}_Q)$ can be constructed as follows
\begin{align}
    \rho_A(\tilde{\mu}_Q) &= \frac{e^{-K_{\text{gen}}}}{\text{Tr}(e^{-K_{\text{gen}}})} = \rho_A \frac{e^{\tilde{\mu}_Q \tilde{Q}_A/\tilde{T}_0}}{n_A(\tilde{\mu}_Q)}.
\end{align}
Here, the normalization factor $n_A(\tilde{\mu}_Q) = \text{Tr}[\rho_A e^{\tilde{\mu}_Q \tilde{Q}_A/\tilde{T}_0}]$ is introduced to ensure that the new density matrix satisfies the unit trace condition, $\text{Tr}[\rho_A(\tilde{\mu}_Q)] = 1$.
The charge R\'enyi entropy is then given by \cite{Belin:2013uta}
\begin{align}
    S_n(\tilde{\mu}_Q) &= \frac{1}{1-n}\log \text{Tr}\left[ \rho_A \frac{e^{\tilde{\mu}_Q \tilde{Q}_A/\tilde{T}_0}}{n_A(\tilde{\mu}_Q)} \right]^n. \label{eq:charged_renyi}
\end{align}
Consequently, the inclusion of the charge sector exponentially reweights the eigenvalues of the reduced density matrix according to their associated local charges. In particular, by keeping the R\'enyi index $n$ fixed while increasing the chemical potential $\tilde{\mu}_Q$, eigenstates carrying larger charges receive an exponential enhancement in their spectral weight. 
As a result, these charged sectors can dominate the generalized reduced density matrix even when their contributions to the original spectrum are subleading. 
The charged R\'enyi entropy at finite $n$ therefore becomes sensitive to the large-eigenvalue or vacuum-like sector of the entanglement spectrum, a regime conventionally associated with the large-$n$ limit. 
In this sense, the chemical potential acts as an additional parameter for reorganizing spectral dominance without requiring an asymptotically large R\'enyi index.

\newpage

\section{Central Charge R\'enyi Entropy}\label{sec:centralRenyi}
For the remainder of this section, we consider thermal CFTs in a fixed $(\tilde{Q},\mathcal{V},\mu_C)$ ensemble on $\mathbb{R}\times \mathbb{H}^{d-1}$ and employ a conformal transformation to derive the corresponding R\'enyi entropy of the reduced density matrix associated with a spherical entangling surface. 
This construction provides a systematic framework for defining and computing a new class of R\'enyi entropies in flat spacetime by incorporating the extended thermodynamic phase space of thermal CFTs.

The thermal density matrix $\rho_\text{therm}$ associated to the thermal CFTs in a fixed $(\tilde{Q},\mathcal{V},\mu_C)$ ensemble can be written as follows
\begin{align}
    \rho_\text{therm}=\frac{e^{-(H-\mu_CC)/\tilde{T}_0}}{Z(\tilde{T}_0,\mu_C)}.\label{eq:thermden}
\end{align}
Here, \(H\) denotes the Hamiltonian generating translations along the thermal circle on \(\mathbb{R}\times \mathbb{H}^{d-1}\) and the partition function takes the form
\begin{align}
    Z(\tilde{T}_0,\mu_C)=\mathrm{Tr}\left[e^{-(H-\mu_C C)/\tilde{T}_0}\right].
\end{align}
It is crucial to emphasize that the evaluation of Eq.~\eqref{eq:thermden} is non-trivial.
If $C$ were interpreted as a fixed $c$-number characterizing a single CFT, the exponential $e^{\mu_C C/\tilde T_0}$ in $Z(\tilde{T}_0,\mu_C)$ would trivially factor out of the trace and cancel against an identical factor in the numerator, rendering $\rho_\text{therm}$ independent of $\mu_C$.
However, in our construction, the trace operation must be formally summed over a family of distinct CFTs with varying central charges. 
Consequently, the density matrix \(\rho_{\mathrm{therm}}\) defines an effective thermal ensemble over theories with varying numbers of degrees of freedom.

Through the CHM map, the conformal transformation between flat space and $\mathbb{R}\times \mathbb{H}^{d-1}$ is implemented by a unitary transformation $U$.
Accordingly, the thermal density matrix $\rho_\text{therm}$ is related to the reduced density matrix $\rho_A$ through
\begin{align}
    \rho_A(\mu_C)=U\rho_\text{therm}(\mu_C)U^{-1}.
\end{align}
From $\rho_\text{therm}(\mu_C)$ in Eq.\eqref{eq:thermden}, we obtain
\begin{align}
    \rho_A(\mu_C) &= \rho_A \frac{e^{\mu_C C/\tilde{T}_0}}{n_A(\mu_C)},
\end{align}
where $\rho_A$ is the ordinary reduced density matrix as defined in Eq.\eqref{eq:Hmod} associated with the modular Hamiltonian $K_A=2\pi R\,H$, and $n_A(\mu_C)=\mathrm{Tr}\left(\rho_A e^{\mu_C C/\tilde{T}_0}\right)$ is the normalization factor.
In this sense, $\rho_A(\mu_C)$ should be viewed as a density matrix on the space of theories rather than a conventional object defined within a single theory.
The R\'enyi entropy can be defined via $\rho_A(\mu_C)$ as follows
\begin{align}
    S_n(\mu_C)= \frac{1}{1-n}\log \text{Tr}\left[ \rho_A \frac{e^{\mu_C C/\tilde{T}_0}}{n_A(\mu_C)} \right]^n. \label{eq:Renyifixmuc}
\end{align}
As a result, Eq.\eqref{eq:Renyifixmuc} admits the following thermal representation
\begin{eqnarray}
    S_n(\mu_C)&=&\frac{1}{1-n}\log \frac{Z(\tilde{T}_0/n,\mu_C)}{Z(\tilde{T}_0,\mu_C)^n}
    =\frac{n}{n-1} \frac{1}{\tilde{T}_0} \left[ \tilde{\Phi}_2(\tilde{T}_0/n,\mu_C)-\tilde{\Phi}_2(\tilde{T}_0,\mu_C) \right], \label{eq:Snmucfromdensity}
\end{eqnarray}
where $\tilde{\Phi}_2$ is the free energy as illustrated in Eq.\eqref{F2 charge}.
We shall refer to this new type of R\'enyi entropy as the \textit{central-charge R\'enyi entropy}.

Having established the central-charge R\'enyi entropy, we now investigate its dependence on the index $n$ and the central charge potential $\mu_C$.
Since the central charge $C$ can be varied with constant $\mu_C$ in this ensemble, we thereby substitute $\tilde{S}_\text{th}$ as expressed in Eq.~\eqref{S3} into Eq.~\eqref{Renyi 2}, which yields
\begin{align}
    S_n(\mu_C)&=\frac{n}{n-1}\frac{1}{\tilde{T}_0}\int_{x_n}^{x_1} dx \frac{2\sqrt{2}\pi \tilde{Q}x^{\frac{d}{2}+1}}{\ell_*\sqrt{(d-1)(d-2)}\sqrt{-x^d-x^{d+2}-\mu_C Rx^2}}\partial_x\tilde{T}(x,\mu_C) \nonumber \\
    &=\frac{n}{n-1} \frac{1}{\tilde{T}_0} \left[ \tilde{\Phi}_2(x_n,\mu_C)-\tilde{\Phi}_2(x_1,\mu_C) \right]. \label{Sn muC}
\end{align}
Note that $\tilde{T}(x,\mu_C)$ defined in Eq.~\eqref{T3} was used to obtain the above result.
Here, $x_n$ is determined by the largest solution of the relation $\tilde{T}(x_n,\mu_C)=\tilde{T_0}/n$, which leads to
\begin{eqnarray}
    x_n^d-\frac{1}{n(d-1)}x_n^{d-1}+\frac{d-2}{2(d-1)}\mu_CR=0. \label{xn sol}   
\end{eqnarray}

\begin{figure}[h]
    \centering
    \includegraphics[width = 7 cm]{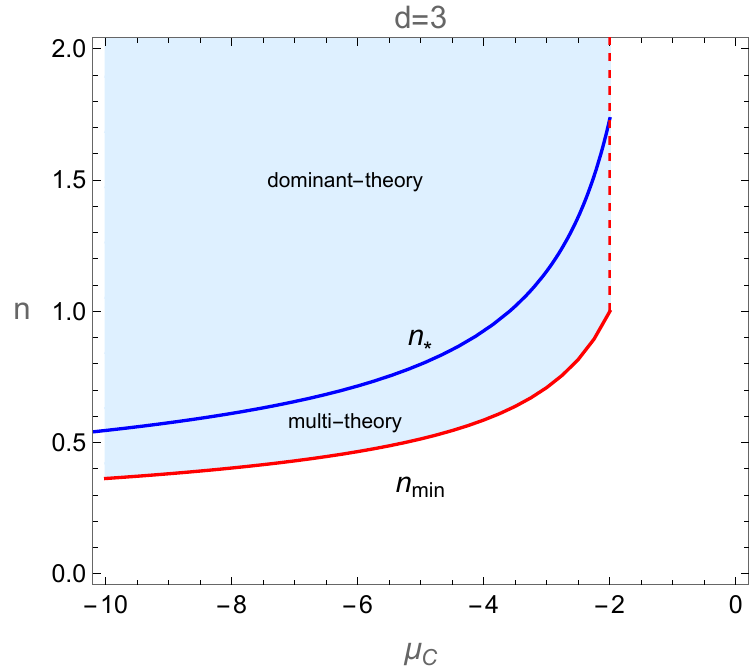}
    \caption{R\'enyi index space for central-charge R\'enyi entropy with $R=1$.} 
    \label{fig: regionn vs muC}
\end{figure}

We consider the case $d=3$, where the $4$-dimensional hyperbolic RN-AdS BH in the bulk is holographically dual to the $3$-dimensional CFT at the boundary.
Since the Gibbs free energy $\tilde{\Phi}_2$ expressed in Eq.~\eqref{Sn muC} is a real-valued function, we observed a lower bound on $n$ and $\mu_C$.
Namely, for a fixed value of $\mu_C$, the condition $\tilde{\Phi}(x_n,\mu_C) \in \mathbb{R}$ sets a lower bound on the R\'enyi index $n_\text{min}$, below which $\tilde{\Phi}_2(x_n,\mu_C)$ becomes imaginary number.
However, the analytic expression for $n_\text{min}$ is rather complicated, we instead plot its numerical values as a function of $\mu_C$, represented by the solid red curve in Fig.~\ref{fig: regionn vs muC}.
Conversely, when varying $\mu_C$ at a fixed $n$, there exists a maximum central charge potential $\mu_{C\,\text{max}} = -2/R$, corresponding to the massless configuration of hyperbolic black hole in the bulk. 
Below this threshold, $S_n$ is undefined as $\tilde{\Phi}_2(x_n,\mu_C)$ becomes complex. 
This boundary is marked by the dashed red vertical line in Fig.~\ref{fig: regionn vs muC}.

Before proceeding further, it is essential to clarify how the eigenvalues of $\rho_A(\mu_C)$ are interpreted in our setup.
The eigenvalues $\lambda$ of $\rho_A(\mu_C)$ admit a generalized interpretation that extends the notion of quantum mechanical probabilities. 
In the present framework, they encode a joint probability distribution spanning both the \textit{state space} and the \textit{theory space}. 
Specifically, each eigenvalue characterizes the probability weight associated not only with a particular configuration of modular energy but also with distinct CFT labeled by their respective central charges.
In this way, the large-$n$ limit of $S_n(\mu_C)$ selectively captures the largest eigenvalues of $\rho_A(\mu_C)$.
This regime corresponds to configurations with low modular energy and the most probable CFT realizations.
In contrast, as $n \rightarrow 0$, $S_n(\mu_C)$ becomes increasingly sensitive to the vast configurations of smaller eigenvalues. 
These characterize highly excited modular energies coupled with rare CFT realizations.
In some sense, $S_n$ acts as a spectral filter.
In the subsequent analysis, we will explicitly determine which specific values of the central charge govern each of these respective regimes.

Motivated by this picture, investigating the index $n_*$ in this extended ensemble becomes particularly compelling.
We propose that $n_*$ serves as a threshold separating two qualitatively distinct statistical regimes in the theory space, namely a \textit{dominant-theory regime} for $n>n_*$, and a \textit{multi-theory regime} for $n<n_*$.
Using Eq.~\eqref{T* 2nd en}, the characteristic index $n_*$ for this ensemble can be defined as follows
\begin{align}
    n_*=\frac{\tilde{T}_0}{\tilde{T}_*}=\frac{\sqrt{2}\pi\left[d(d-1)x_\text{min}^2-(d-2)\right]}{\ell_*d\left[(d-1)(dx_\text{min}^2-(d-2))\right]^{3/2}}. \label{eq:ns2}
\end{align}
The solid blue curve in Fig.~\ref{fig: regionn vs muC} illustrates $n_*$ as a function of $\mu_C$, which increases monotonically with $\mu_C$.
Since $\mu_C$ is bounded by $\mu_{C\,\text{max}}$, the characteristic index also attains a maximum value $n_*=1.734$.
Therefore, any central-charge R\'enyi entropy with $n>1.734$ is  always dominated by the dominant-theory regime. 
In the parameter region above the $n_*(\mu_C)$ curve, the central-charge R\'enyi entropy primarily probes entanglement contributions from the dominant-theory regime of the reduced density matrix. 
In contrast, in the region below this curve, the central-charge R\'enyi entropy becomes sensitive to the multi-theory regime. 
\begin{figure}[h]
    \centering
    \includegraphics[width = 7 cm]{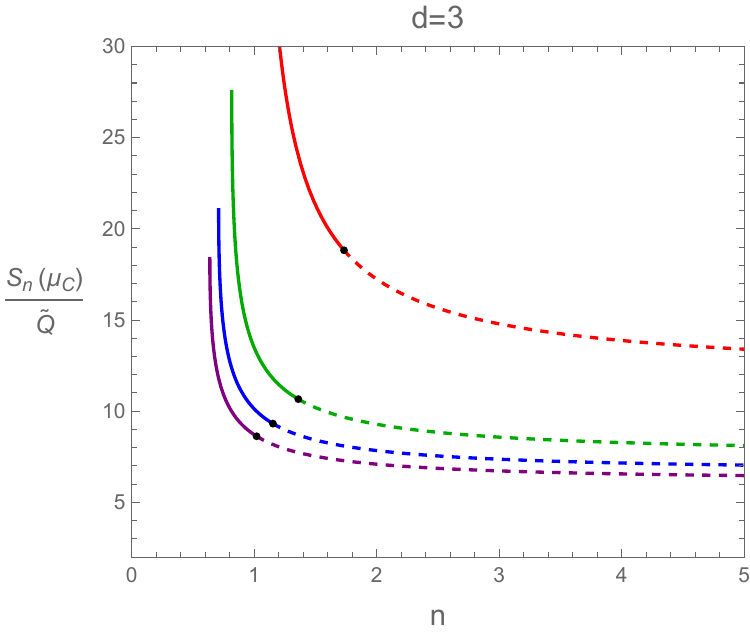} \hspace{1 cm}
    \includegraphics[width = 7 cm]{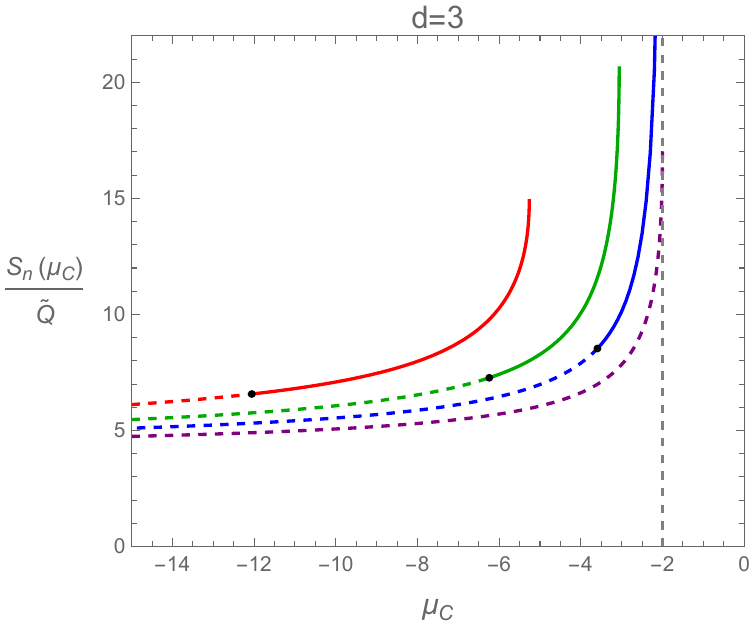}
\caption{\textbf{Left:} The $d=3$ R\'enyi entropy normalized by charge $S_n(\mu_C)/\tilde{Q}$ as a function of $n$ is plotted with $\mu_C=-2.0$ (red), $-2.5$ (green), $-3.0$ (blue) and $-3.5$ (purple). \textbf{Right:} The $S_n(\mu_C)/\tilde{Q}$ versus $\mu_C$ with $n=0.50$ (red), $0.70$ (green), $1.0$ (blue) and $2.0$ (purple). Note that we use $R=1$ and $\ell_*=1$ in these plots.}  
    \label{fig:2ndSnd=3}
\end{figure}

We illustrate the results for R\'enyi entropy normalized by charge $S_n(\mu_C)/\tilde{Q}$ as a function of R\'enyi index $n$ for various values of $\mu_C$ in the left panel of Fig.~\ref{fig:2ndSnd=3}.
For $n \geq n_\text{min}$, the ratio $S_n(\mu_C)/\tilde{Q}$ reaches its maximum value at $n_\text{min}$ and decreases monotonically as $n$ increases.
For the right panel in Fig.~\ref{fig:2ndSnd=3}, we illustrate how the R\'enyi entropy of order $n$ varies with the central charge potential $\mu_C$.
The results indicate that $S_n(\mu_C)/\tilde{Q}$ has a maximum value at $\mu_{C\,\text{max}}$ and decreases monotonically when $\mu_C$ has more negative value.
Unlike the others values of $n$, the maximum value of $S_n(\mu_C)/\tilde{Q}$ at $n=1$ is infinite, when $\mu_C\to \mu_{C\,\text{max}}$ as shown in the blue curve in the right panel of Fig.~\ref{fig:2ndSnd=3}.
Since the probability weight of a distinct CFT contributing to the ensemble scales with the factor $e^{\mu_C C/\tilde{T}_0}$, CFTs with large central charges are exponentially suppressed when $\mu_C$ is highly negative.
As a result, the generalized ensemble is restricted to a narrower theory space governed by lower degrees of freedom.
Consequently, the central-charge R\'enyi entropy of order $n$ remains relatively small due to the restricted amount of available degrees of freedom contributing to the entanglement.

As in the previous ensemble, the black dots in the left panel of Fig.~\ref{fig:2ndSnd=3} divide the curves into two regions of $n<n_*$ (solid curves) and $n>n_*$ (dashed) curves.
We find that as $\mu_C$ becomes increasingly negative, the characteristic R\'enyi index $n_*$ shifts toward smaller values. 
Since the R\'enyi entropy of order $n>n_*$ is dominated by the dominant-theory regime, this shift implies that an increasingly larger range of R\'enyi indices probes that regime.

\begin{figure}[h]
\centering
\includegraphics[width = 7 cm]{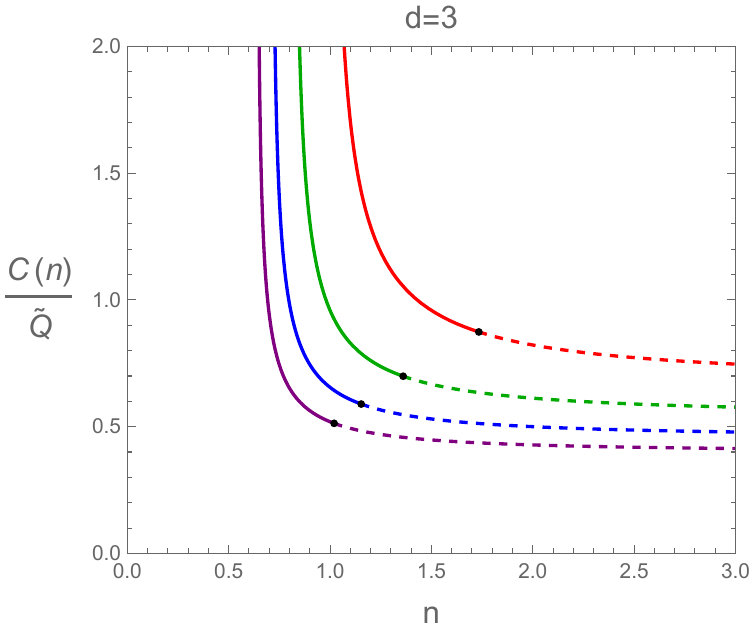}
\caption{The $d=3$ effective central charge $C(n)/\tilde{Q}$ as a function of $n$ is plotted with $\mu_C=-2.0$ (red), $-2.5$ (green), $-3.0$ (blue) and $-3.5$ (purple).} 
\label{fig:2nd_ns_behavior}
\end{figure}

To investigate the effect of $\mu_C$ on the central-charge R\'enyi entropy of order $n$, we introduce the characteristic central charge potential $\mu_{C*}$, which  separates the dominant and multi-theory regimes.
For a given R\'enyi index $n$, the corresponding value of $\mu_{C*}$ can be obtained from Eq.\eqref{eq:ns2}, as indicated by the black dots in the right panel of Fig.~\ref{fig:2ndSnd=3}.
As $n$ increases, $\mu_{C*}$ also increases, implying that a larger portion of the $\mu_C$ is occupied by the dominant-theory regime.
Note that for $n=2>1.734$, the central-charge R\'enyi entropy is entirely dominated by the dominant-theory regime, as illustrated by the dashed-purple curve.
The behavior of $S_n(\mu_C)/\tilde{Q}$ at $n=n_*$ as a function of $\mu_C$ is displayed in Fig.\ref{fig:2nd_ns_behavior}.

We now proceed to determine the values of the central charge within each statistical regime by analyzing the central charge $C$ as a function of the index $n$ with different value of $\mu_C$, as shown in Fig.~\ref{fig:2nd_ns_behavior}.
In the small-$n$ limit ($n < n_*$), which characterizes the multi-theory regime, the effective central charge $C(n)/\tilde{Q}$ is significantly enhanced and reaches its maximum at $n=n_\text{min}$.
This regime corresponds to rare CFT realizations with large central charges and highly excited modular energies, whose statistical weights are exponentially suppressed by negative $\mu_C$.
Conversely, as $n$ increases and enters the dominant-theory regime ($n > n_*$), the effective central charge drops and approach to some specific value. 
In this large-$n$ domain, the index $n$ isolates the largest eigenvalues of the reduced density matrix, meaning the ensemble is strictly dominated by the most probable CFT realizations with lower degrees of freedom.

\begin{figure}[h]
    \centering
    \includegraphics[width = 6 cm]{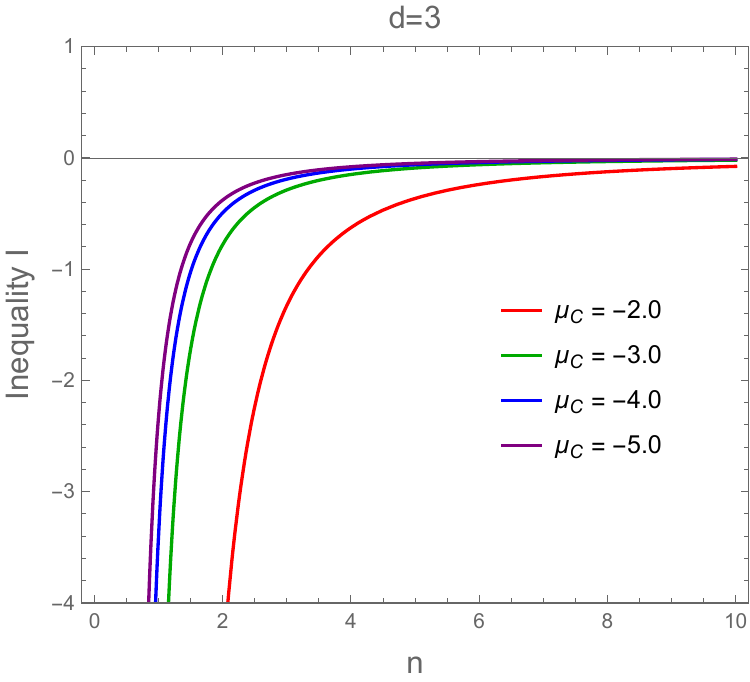} \hspace{1 cm}
    \includegraphics[width = 6 cm]{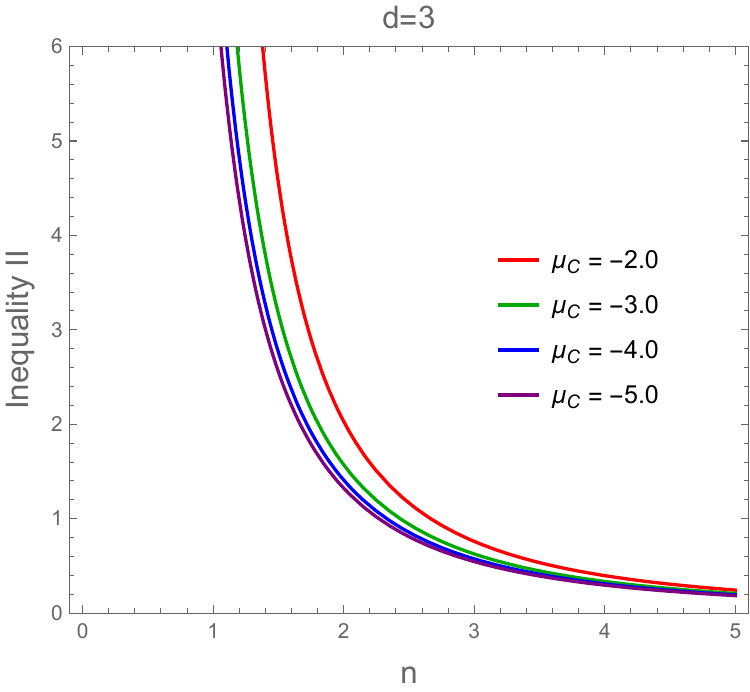} \\
    \includegraphics[width = 6 cm]{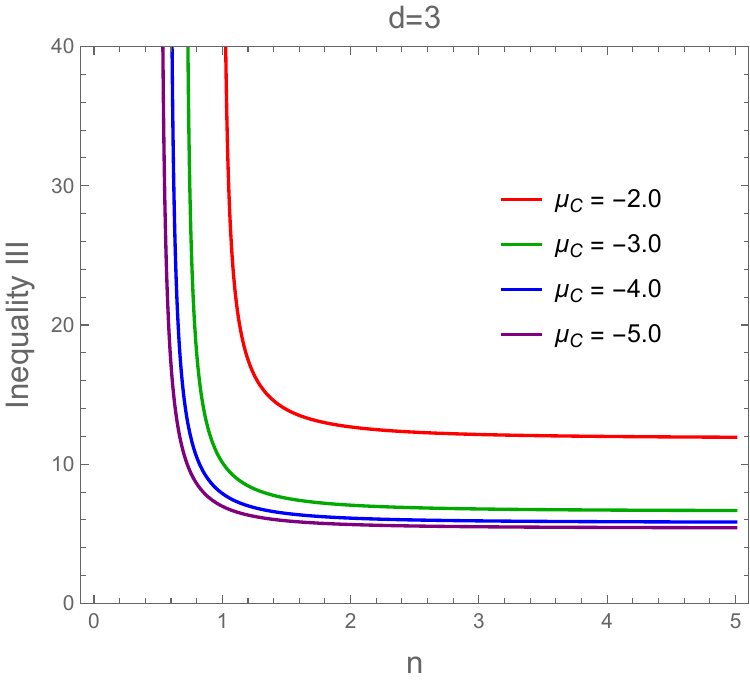} \hspace{1 cm}
    \includegraphics[width = 6 cm]{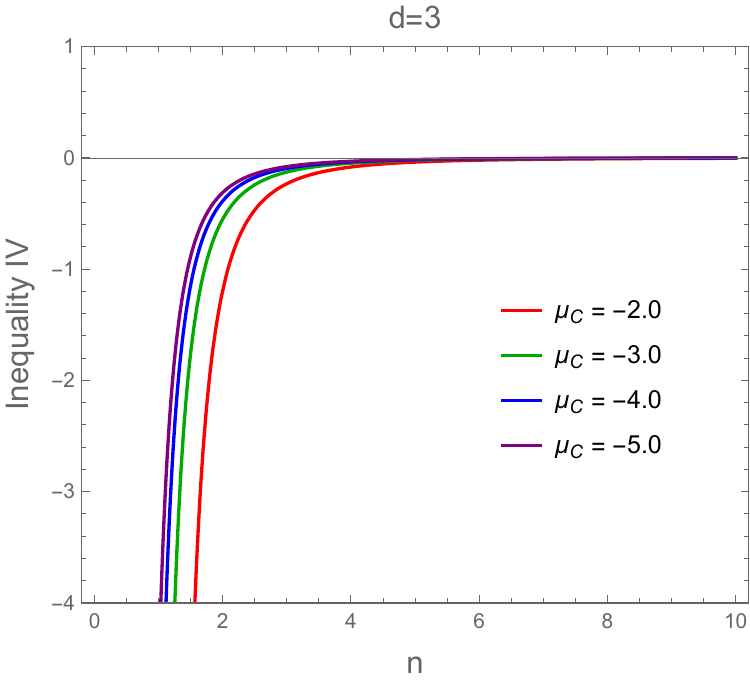}
\caption{
The derivative expressions $\displaystyle \frac{\partial S_n}{\partial n},\, \frac{\partial}{\partial n}\left(\frac{n-1}{n}S_n\right),\, \frac{\partial}{\partial n}\left( (n-1)S_n \right)$ and $\displaystyle \frac{\partial^2}{\partial n^2}\left((n-1)S_n\right)$ (labeled as Inequality I, II, III, and IV, respectively) for the central-charge Rényi entropy as a function of $n$ are plotted with different values of $\mu_C$.
}
\label{fig:inequalities}
\end{figure}

To ensure the physical validity of our extended framework, any well-defined R\'enyi entropy must be non-negative and strictly satisfy the four inequalities described in Eqs.~\eqref{eq:I}-\eqref{eq:IV}. 
In Fig.~\ref{fig:inequalities}, we evaluate the derivatives, namely $\displaystyle \frac{\partial S_n}{\partial n},\, \frac{\partial}{\partial n}\left(\frac{n-1}{n}S_n\right),\, \frac{\partial}{\partial n} \left((n-1)S_n \right)$ and $\displaystyle \frac{\partial^2}{\partial n^2}\left((n-1)S_n\right)$ for the central-charge R\'enyi entropy as a function of $n$ with different values of $\mu_C$.
From a thermodynamic perspective of dual CFT in fixed $(\tilde{Q},\mathcal{V},\nu_C)$ ensemble, the first and fourth inequalities correspond to the positivity of the heat capacity, whereas the second and third inequalities guarantee a positive thermal entropy, as confirmed by our results in Fig.~\ref{fig: 3rd ensemble thermo}.

\section{Conclusion and Outlook}\label{sec:concl}

In this work, we have introduced and investigated the central-charge R\'enyi entropy as given in Eq.~\eqref{eq:Renyifixmuc}, which generalizes the conventional R\'enyi entropy by incorporating the central charge potential $\mu_C$. 
We interpret this quantity as a measure of entanglement among different theories associated with a statistical ensemble of holographic CFT with fluctuating numbers of degrees of freedom. 
As shown in Sec. [VI], this central-charge R\'enyi entropy was constructed for a CFT with a spherical entangling surface through the conformal thermodynamic framework and the CHM map, which relates the thermal density matrix $\rho_\text{therm}$ of a CFT on a hyperbolic cylinder in the fixed $(\tilde{Q},\mathcal{V},\mu_C)$ ensemble to the corresponding reduced density matrix $\rho_A$ associated with the spherical region.

To establish a systematic investigation of this extended entropy, we comprehensively analyzed the thermal properties of the dual CFT on $\mathbb{R} \times \mathbb{H}^{d-1}$—holographically dual to hyperbolic RN-AdS BH—using the conformal thermodynamics framework. 
Our analysis revealed a critical dimension of $d=4$ in the $p-R$ plane of fixed $(\tilde{\mu}_Q,\mathcal{V},C)$ ensemble. 
In lower dimensions ($d=2,3$), the CFT possesses a single mechanically unstable branch with $\kappa < 0$, whereas for $d \geq 4$, two distinct branches emerge: a mechanically unstable branch $\kappa < 0$ at small $R$, and a stable one $\kappa > 0$ at large $R$.
Intriguingly, regardless of mechanical instability, all branches remain thermally stable above extremality as indicated by their positive heat capacity.

For the fixed $(\tilde{Q},\mathcal{V},\mu_C)$ ensemble, the mechanical property of dual CFT becomes richer, exhibiting three distinct branches in the $p-R$ plane: an unstable branch at small $R$, a stable branch at intermediate $R$, and another unstable branch at large $R$.
However, we did not observe van der Waals-like behavior, suggesting that central charge fluctuations affect the mechanical phase structure without inducing a criticality. 
Nevertheless, all three branches remain thermally stable as a result of their positive heat capacity.
Thus, the dual CFT of the intermediate $R$ branch exhibits simultaneous mechanical and thermal stability.

Beyond novel mechanical phase structures of the dual CFT, conformal thermodynamics also provides a framework to revisit the long-standing puzzle of residual entropy in extremal hyperbolic BHs.  
Previous studies interpreted the residual entropy as evidence of a highly degenerate ground state, despite the absence of supersymmetry. 
From the generalized Euler relation, our analysis suggests a more refined thermodynamic interpretation: the residual entropy does not arise from thermal excitations but is instead associated with the central-charge sector, which encodes the number of degrees of freedom of the CFT. 
The separation between this central-charge contribution and thermal excitations naturally gives rise to a thermodynamic mass gap $M_\text{gap}$, which defines the characteristic temperature scale $T_*$.
In contrast in the conventional thermal ensemble, our analysis in the fixed $(\tilde{Q},\mathcal{V},\mu_C)$ ensemble reveals that the residual entropy and the corresponding residual central charge are controlled by the central charge potential $\mu_C$.

Translating this thermodynamic insight through the CHM map, we relate $T_*$ to the characteristic R\'enyi index $n_*=\tilde{T}_0/\tilde{T}_*$ (as detailed in Sec.~\ref{sec:index}), which separates two qualitatively different sectors of the entanglement spectrum.
For charged R\'enyi entropy, when $n>n_*$ the R\'enyi entropy predominantly probes the largest eigenvalues of the reduced density matrix, corresponding to a vacuum-like sector associated with the near-extremal regime of the dual CFT. 
In contrast, the regime $n<n_*$ becomes increasingly sensitive to smaller eigenvalues, where thermal-like modular excitations become important.

In Sec.~\ref{sec:centralRenyi}, we construct the central-charge R\'enyi entropy and explore the interpretation of its characteristic R\'enyi index $n_*$. 
Unlike the charged R\'enyi entropy, where $n_*$ separates the vacuum-like and thermal-like sectors of the modular spectrum, $n_*$ within the central-charge R\'enyi entropy distinguishes the crossover between the multi-theory and dominant-theory regimes. 
Specifically, the regime $n>n_*$ is dominated by a single CFT with a particular central charge, while the regime $n<n_*$ receives significant contributions from multiple CFTs with different numbers of degrees of freedom. 
This provides a new perspective on how the entanglement structure evolves across a family of holographic CFTs connected through variations of the central charge.

Finally, it would be interesting to extend our analysis to more general gravitational solutions and investigate quantum corrections near extremality through JT gravity \cite{Iliesiu:2020qvm}. 
Such directions may provide a deeper understanding of the universality of the mass gap and the underlying quantum structure encoded in holographic R\'enyi entropy.

\section*{Acknowledgement}

The authors thank Weerawit Horinouchi for valuable discussions and insights during the development of this work. This research was supported by King Mongkut's University of Technology Thonburi (KMUTT), Thailand Science Research and Innovation (TSRI), and the National Science, Research and Innovation Fund (NSRF) Fiscal year 2024 [grant number FRB670073/0164]. This research has also received funding support from the NSRF via the Program Management Unit for Human Resources \& Institutional Development, Research and Innovation [grant number B13F670070]. 

\newpage

\appendix

\section{Derivation of CFT Thermodynamics}\label{App A}
The mass of hyperbolic charged AdS-BH in Eq.~\eqref{charged BH mass} can be expressed in terms of the area at the horizon $A_H=V_{\mathbb{H}^{d-1}}\rho_+^{d-1}$, the electric charge $Q$ and the cosmological constant $\Lambda$ as follows
\begin{eqnarray}
    M(A_H,Q,\Lambda)=-\frac{A_H^{\frac{d}{d-1}}}{8\pi G_\text{N}V_{\mathbb{H}^{d-1}}^{\frac{1}{d-1}}}\frac{\Lambda}{d}-\frac{(d-1)V_{\mathbb{H}^{d-1}}^{\frac{1}{d-1}}A_H^{\frac{d-2}{d-1}}}{16\pi G_\text{N}}+\frac{8\pi G_\text{N}}{(d-2)V_{\mathbb{H}^{d-1}}^{\frac{1}{d-1}}}\frac{Q^2}{A_H^{\frac{d-2}{d-1}}\ell_*^2}, \label{Mass for Smarr}
\end{eqnarray}
where $\Lambda=-d(d-1)/(2L^2)$.

This mass is also found to satisfy $M(\alpha^{d-1}A_H,\alpha^{d-2}Q,\alpha^{-2}\Lambda)=\alpha^{d-2}M(A_H,Q,\Lambda)$ where $\alpha$ is a nonzero scalar.
This implies that the mass $M\big(A_H^{1/(d-1)}, Q^{1/(d-2)}, \Lambda^{-1/2}\,\big)$ is said to be a homogeneous function of degree $d-2$.
From the Euler's theorem, it satisfies the algebraic relation:
\begin{eqnarray}
    (d-2)M=(d-1)\frac{\partial M}{\partial A_H}A_H+(d-2)\frac{\partial M}{\partial Q}Q+(-2)\frac{\partial M}{\partial \Lambda}\Lambda.
\end{eqnarray}
The partial derivative term in the above equation can be derived from Eq.~\eqref{Mass for Smarr} as
\begin{eqnarray}
    \frac{\partial M}{\partial A_H}A_H=\frac{\partial M}{\partial S_\text{th}}S_\text{th}=TS_\text{th},\qquad \frac{\partial M}{\partial Q}Q=\frac{\mu_Q}{2\pi }Q,\qquad \frac{\partial M}{\partial \Lambda}\Lambda =\frac{\Theta \Lambda}{8\pi G_\text{N}},
\end{eqnarray}
where $\Theta$ is a conjugate thermodynamic variable of $\Lambda$.
The mass formula is then
\begin{eqnarray}
    M=\left(\frac{d-1}{d-2}\right)TS_\text{th}+\frac{\mu_Q}{2\pi}Q-\left(\frac{2}{d-2}\right)\frac{\Theta \Lambda}{8\pi G_\text{N}}. \label{pre Smarr}
\end{eqnarray}
In the extended phase space approach, we identify a bulk pressure and thermodynamic volume as follows:
\begin{eqnarray}
    P=-\frac{\Lambda}{8\pi G_\text{N}} \ \ \ \text{and} \ \ \ V=-\Theta=\frac{V_{\mathbb{H}^{d-1}}}{d}\rho_+^d.
\end{eqnarray}
Substituting them into Eq.~\eqref{pre Smarr}, the generalized Smarr formula and its corresponding first law are obtained as expressed in Eqs.~\eqref{Smarr eps} and \eqref{1st law eps}.

By treating the conformal factor $\omega$ as a thermodynamic variable in the extended conformal thermodynamics framework, the variation of rescaling mass is given by 
\begin{eqnarray}
    d\left( \frac{M}{\omega}\right)=\frac{1}{\omega}dM-\frac{M}{\omega^2}d\omega. \label{diff rescale mass}
\end{eqnarray}
With Eqs.~\eqref{Smarr eps} and \eqref{1st law eps}, the variation of mass $dM$ as given in \cite{Visser:2021eqk,Cong:2021jgb} becomes
\begin{eqnarray}
    dM&=&Td\left(\frac{A_H}{4G_\text{N}}\right)+\frac{\mu_Q}{2\pi L}d(QL)-\frac{M}{d-1}\frac{dL^{d-1}}{L^{d-1}} \nonumber \\
    &&\hspace{2.2 cm} +\left(M-TS_\text{th}-\frac{\mu_Q}{2\pi L}QL\right)\frac{d(L^{d-1}/G_\text{N})}{L^{d-1}/G_\text{N}}. \label{dM holographic thermo}
\end{eqnarray}
Substituting Eq.~\eqref{dM holographic thermo} into Eq.~\eqref{diff rescale mass}, we obtain
\begin{eqnarray}
    d\left(\frac{M}{\omega}\right)&=&\frac{T}{\omega}d\left(\frac{A_H}{4G_\text{N}}\right)+\frac{\mu_Q} {2\pi \omega L}d(QL)-\frac{M}{(d-1)\omega}\frac{d(\omega L)^{d-1}}{(\omega L)^{d-1}}\nonumber \\
    && \hspace{2.3cm}+\left( \frac{M}{\omega}-\frac{TS_\text{th}}{\omega}-\frac{\mu_Q Q}{2\pi \omega}\right)\frac{d(L^{d-1}/G_\text{N})}{L^{d-1}/G_\text{N}}.\label{d scaling M}
\end{eqnarray}
Note that $\displaystyle \frac{M}{(d-1)\omega}\frac{d(\omega L)^{d-1}}{(\omega L)^{d-1}}=\frac{M}{(d-1)\omega}\frac{dL^{d-1}}{L^{d-1}}+\frac{M}{\omega^2}d\omega$.
As proposed in \cite{PhysRevLett.130.181401}, the holographic form for the dual CFT thermodynamic quantities (with tildes) are scaled by the conformal factor $\omega$ of the bulk thermodynamic quantities (without tildes) as follows:
\begin{eqnarray}
    \tilde{E}=\frac{M}{\omega},\qquad
    \tilde{S}_\text{th}=S_\text{th},\qquad
    \tilde{T}=\frac{T}{\omega},\qquad
    \tilde{Q}=Q L,\qquad
    \tilde{\mu}_Q=\frac{\mu_Q}{\omega L}. \label{dictionary}
\end{eqnarray}
Consequently, Eq.~\eqref{d scaling M} can be written in the CFT's thermodynamic quantities as
\begin{eqnarray}
d\tilde{E}=\tilde{T}d\tilde{S}_\text{th}+\frac{\tilde{\mu}_Q}{2\pi}d\tilde{Q}-pd\mathcal{V}+\mu_CdC,\label{1st law CFT}
\end{eqnarray}
where
\begin{eqnarray}
    \mu_C&=&\frac{1}{C}\left(\tilde{E}-\tilde{T}\tilde{S}_\text{th}-\frac{\tilde{\mu}_Q}{2\pi}\tilde{Q}\right),\qquad \label{Euler} \\
    p&=&\frac{\tilde{E}}{(d-1)\mathcal{V}}. \label{p}
\end{eqnarray}

\bibliography{ref}

\end{document}